\documentclass[preprint,12pt,eqsecnum,nofootinbib,amsmath,amssymb]{revtex4}

\newcommand{\datechange}{Jan 11 2019}

\newcommand{\laurnumber}{LA-UR-19-20366}

\newcommand{\mytitle}{Floating-Point Calculations on a Quantum Annealer: \\
Division and Matrix Inversion
}
\usepackage{graphicx}  
\usepackage{dcolumn}  
\usepackage{bm}          
\usepackage{latexsym}  
\usepackage{fancyhdr}  
\usepackage{wrapfig}
\usepackage{comment}
\usepackage{dsfont}
\usepackage{mathtools}
\usepackage{datetime}

\newcommand{\smN}{{\rm\scriptscriptstyle N}}
\newcommand{\bvec}[1]{\mathbf{#1}}

\newcommand{\smT}{{\rm\scriptscriptstyle T}}

\newcommand{\smR}{{\rm\scriptscriptstyle R}}
\newcommand{\smCT}{{\rm\scriptscriptstyle CT}}

\newcommand{\bodyskip}{\baselineskip 18pt plus 1pt minus 1pt}
\newcommand{\bibskip}{\baselineskip16pt plus 1pt minus 1pt}
\newcommand{\tableofcontentsskip}{\baselineskip 14pt plus 1pt minus 1pt}
\newcommand{\footnoteskip}{\baselineskip 12pt plus 1pt minus 1pt}
\newcommand{\abstractskip}{\baselineskip 13pt plus 1pt minus 1pt}
\newcommand{\titleskip}{\baselineskip 18pt plus 1pt minus 1pt}
\newcommand{\affiliationskip}{\baselineskip 15pt plus 1pt minus 1pt}

\def\frame#1#2#3#4{\vbox{\hrule height #1pt    
  \hbox{\vrule width #1pt\kern #2pt                     
  \vbox{\kern #2pt                                               
  \vbox{\hsize #3\noindent #4}                            
  \kern #2pt}                                                        
  \kern #2pt\vrule width #1pt}                              
  \hrule height0pt depth #1pt}                            
}
\def\myframe#1{\vbox{\hrule height 0.1pt    
  \hbox{\vrule width 0.1pt\kern 2pt                     
  \vbox{\kern 2pt                                               
  \vbox{\hsize 16.5cm\noindent #1}                            
  \kern 2pt}                                                        
  \kern 2pt\vrule width 0.1pt}                              
  \hrule height0pt depth 0.1pt}                            
}

\def\fitframe #1#2#3{\vbox{\hrule height#1pt  
  \hbox{\vrule width#1pt\kern #2pt             
  \vbox{\kern #2pt\hbox{#3}\kern #2pt}         
  \kern #2pt\vrule width#1pt}                  
  \hrule height0pt depth#1pt}                  
}
\def\shframe #1#2#3#4{\vbox{\hrule height 0pt 
 \hbox{\vrule width #1pt\kern 0pt             
 \vbox{\kern-#1pt\frame{.3}{#2}{#3}{#4}       
 \kern-.3pt}                                  
 \kern-#2pt\vrule width 0pt}                  
 \hrule height #1pt}                          
}
\begin{document}

%

\hfill \laurnumber
\title{\titleskip
  \mytitle
}

\author{Michael L Rogers and Robert L Singleton Jr}
\vskip 0.2cm 
\affiliation{\affiliationskip
     Los Alamos National Laboratory\\
     Los Alamos, New Mexico 87545, USA
}

\date{\datechange}

\begin{abstract}
\abstractskip
\vskip0.3cm 

Systems of linear equations are employed almost universally across a wide range 
of disciplines, from physics and engineering to biology, chemistry and  statistics. 
Traditional solution methods such as Gaussian elimination become very time 
consuming for large matrices, and more efficient computational methods are 
desired. In the twilight of Moore\rq{}s Law, quantum computing is perhaps the 
most direct path out of the darkness. There are two complementary paradigms 
for quantum computing, namely, gated systems and quantum annealers. In this 
paper, we express floating point operations such as division and matrix inversion 
in terms of a {\em quadratic unconstrained binary optimization} (QUBO) problem, 
a formulation that is ideal for a quantum annealer. We first address floating point 
division, and then move on to matrix inversion. We provide a general algorithm
for any number of dimensions, but we provide results from the D-Wave quantum
anneler for $2\times 2$ and $3 \times 3$ general matrices. Our algorithm 
scales to very large numbers of linear equations. We should also mention that 
our algorithm provides the full solution the the matrix problem, while HHL 
provides only an expectation value. 

\end{abstract}
\maketitle


\pagebreak
\tableofcontentsskip
\tableofcontents

\newpage
\bodyskip

\pagebreak
\clearpage
\section{Introduction}

Systems of linear equations are employed almost universally across a wide range 
of disciplines, from physics and engineering to biology, chemistry and  statistics. 
An interesting physics application is computational fluid dynamics (CFD), which 
requires inverting very large matrices to advance the state of the hydrodynamic 
system from one time step to the next. An application of importance in biology 
and chemistry would include the protein folding problem. For large
matrices, Gaussian elimination and other standard techniques becomes 
too time consuming, and therefore faster computational methods are desired. 
As Moore\rq{}s Law draws to a close,  quantum computing offers the most direct 
path forward, and perhaps the most radical path.  In a nutshell, quantum computers 
are physical systems that exploit the laws of quantum mechanics to perform 
arithmetic and logical operations exponentially faster than a conventional computer.
In the words of Harrow, Hassidim, and Lloyd (HHL)\,\cite{hhl}, ``quantum computers 
are devices that harness quantum mechanics to perform computations in ways that 
classical computers cannot.\rq{}\rq{}  There are currently two complementary paradigms 
for quantum computing, namely, gated systems and quantum annealers. Gated systems 
exploit the deeper properties of quantum mechanics such coherence, entanglement 
and non-locality, while quantum annealers take advantage of tunneling between 
metastable states and the ground state. In Ref.~\cite{hhl}, HHL introduces a gated 
method by which the inverse of a matrix can be computed, and 
Refs.~\cite{ref2013a, ref2013b, ref2013c} provide implementations of the algorithm 
to invert $2 \times 2$ matrices. Gated methods are limited by the relatively
small number of qubits that can be entangled into a fully coherent quantum state, 
currently of order $32$ or so. An alternative approach to quantum computing is  the 
quantum anneler\,\cite{fggs}, which takes advantage of quantum tunneling between 
metastable states and the ground state. The D-Wave Quantum Annealers have reached 
capacities of $2000+$ qubits, which suggests that quantum annealers could be quite 
effective for linear algebra with hundreds to thousands of degrees of freedom. In this 
paper, we express floating point operations such as division and matrix inversion as 
{\em quadratic unconstrained binary optimization} (QUBO) problems, which are ideal 
for a quantum annealer. We should mention that our algorithm provides the full
solution the the matrix problem, while HHL provides only an expectation value. 
Furthermore, our algorithm places no contraints on the matrix that we are inverting,
such as a sparcity condition. 

The first step in mapping a general problem to a QUBO problem begins with constructing a 
Hamiltonian that encodes the logical problem in terms of a set of qubits. Next, it will 
be necessary to ``embed'' the problem on  the chip, first by mapping each logical qubit to 
a collection or ``chain'' of physical qubits, and then by determining parameter settings for 
all the physical qubits, including the chain couplings. We have implemented our algorithms 
on the D-Wave 2000Q and 2X chips, illustrating that division and matrix inversion 
can indeed be performed on an existing quantum annealer. The algorithms that we propose
should scale quite well for large numbers of equations, and should be applicable to matrix 
inversion of relatively high order (although probably not exponentially higher order as
in HHL). 

Before examining the various algorithms, it is useful to review the basic formalism 
and to establish some notation. The general problem starts with a graph ${\cal G} 
=( {\cal V} ,  {\cal E} )$, where ${\cal V}$ is the vertex set and ${\cal E}$ is the edge set. 
The QUBO {\em Hamiltonian} on ${\cal G}$ is defined by
\begin{eqnarray}
  H_{\scriptscriptstyle {\cal G}}[Q] 
  &=&
  \sum_{r \in {\cal V}} A_r\, Q_r
  +
  \sum_{rs \in {\cal E}} B_{r s} \,  Q_r  Q_s
  \ ,
  \label{QUBO_1}
\end{eqnarray}
with $Q_r \in \{0, 1 \}$ for all $r \in {\cal V}$. The coefficient $A_r$ is called the
{\em weight} at vertex $r$, while the coefficient $B_{rs}$ is called the {\em strength} 
between vertices $r$ and $s$.  It might be better to call (\ref{QUBO_1}) 
the {\em objective function} rather than the Hamiltonian, as $H_{\scriptscriptstyle {\cal G}}$ 
is a real-valued function and not an operator on a Hilbert-space. However, it is easy to 
map (\ref{QUBO_1}) in an equivalent Hilbert space form, 
\begin{eqnarray}
  \hat H_{\scriptscriptstyle {\cal G}}
  &=&
  \sum_{r \in {\cal V}} A_r\, \hat Q_r
  +
  \sum_{r s \in {\cal E}} B_{r s} \,  \hat Q_r  \hat Q_s
  \ ,
  \label{QUBO_2}
\end{eqnarray}
where $\hat Q_r \vert Q \rangle = Q_r \vert Q \rangle$ for all $r \in {\cal V}$, and 
$\vert Q \rangle \in {\cal H}$ for  Hilbert space $\cal H$. The hat denotes 
an operator on the Hilbert space, and $Q_r$ is the corresponding Eigenvalue of
$\hat Q_r$ with Eigenstate $\vert Q \rangle$. Consequently, we can write
\begin{eqnarray}
  \hat H_{\scriptscriptstyle {\cal G}} \vert Q \rangle 
  = H_{\scriptscriptstyle {\cal G}}[Q] \, \vert Q \rangle
  \ ,
  \label{QUBO_3}
\end{eqnarray}
and we use the terms {\em Hamiltonian} and {\em objective function} interchangeably.
By the {\em QUBO problem}, we mean the problem of finding the lowest energy state 
$\vert Q \rangle $ of the Hamiltonian~(\ref{QUBO_2}), which corresponds to minimizing 
Eq.~(\ref{QUBO_1}) with respect to the $Q_r$. This is an NP-hard problem uniquely 
suited to a quantum annealer.  Rather than sampling all $2^{\# {\cal V} }$ possible 
states, quantum tunneling finds the {\em most likely} path to the ground state by 
minimizing the Euclidian action. In the case of the D-Wave 2X chip, the number of 
distinct quantum states is of order the insanely large number $2^{1000}$, and the 
ground state is selected from this jungle of quantum states by tunneling to those
states with a smaller Euclidean action. 

The Ising model\,\cite{Ising1925} is perhaps the quintessential 
physical example of a QUBO problem, and indeed, is one of the most studied systems in 
statistical physics. The Ising model consists of a square lattice of
\hbox{spin-1/2} particles with 
nearest neighbor spin-spin interactions between sites $r$ and $s$, and when the 
system is immersed in a nonuniform magnetic field, this introduces coupling terms 
at individual sites $r$, thereby producing a Hamiltonian of the form 
\begin{eqnarray}
  H_{\scriptscriptstyle {\cal G}}[J]
  &=&
  \sum_{r \in {\cal V}} B_r\, J_r
  +
  \sum_{rs \in {\cal E}} J_{r s} \,  J_r  J_s
  \ ,
  \label{QUBO_J}
\end{eqnarray}
where $J_r =\pm 1/2$. The Ising problem is connected to the QUBO problem by
$J_r = Q_r - 1/2$. 

\begin{figure}[t!]
\begin{minipage}[c]{0.45\linewidth}
\includegraphics[scale=0.35]{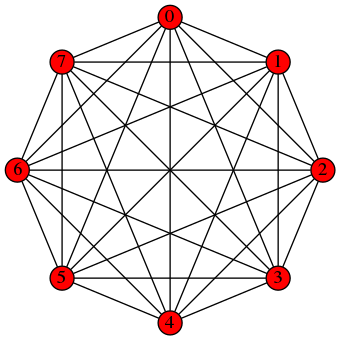} 
\end{minipage}
\hfill
\begin{minipage}[c]{0.5\linewidth}
\includegraphics[scale=0.40]{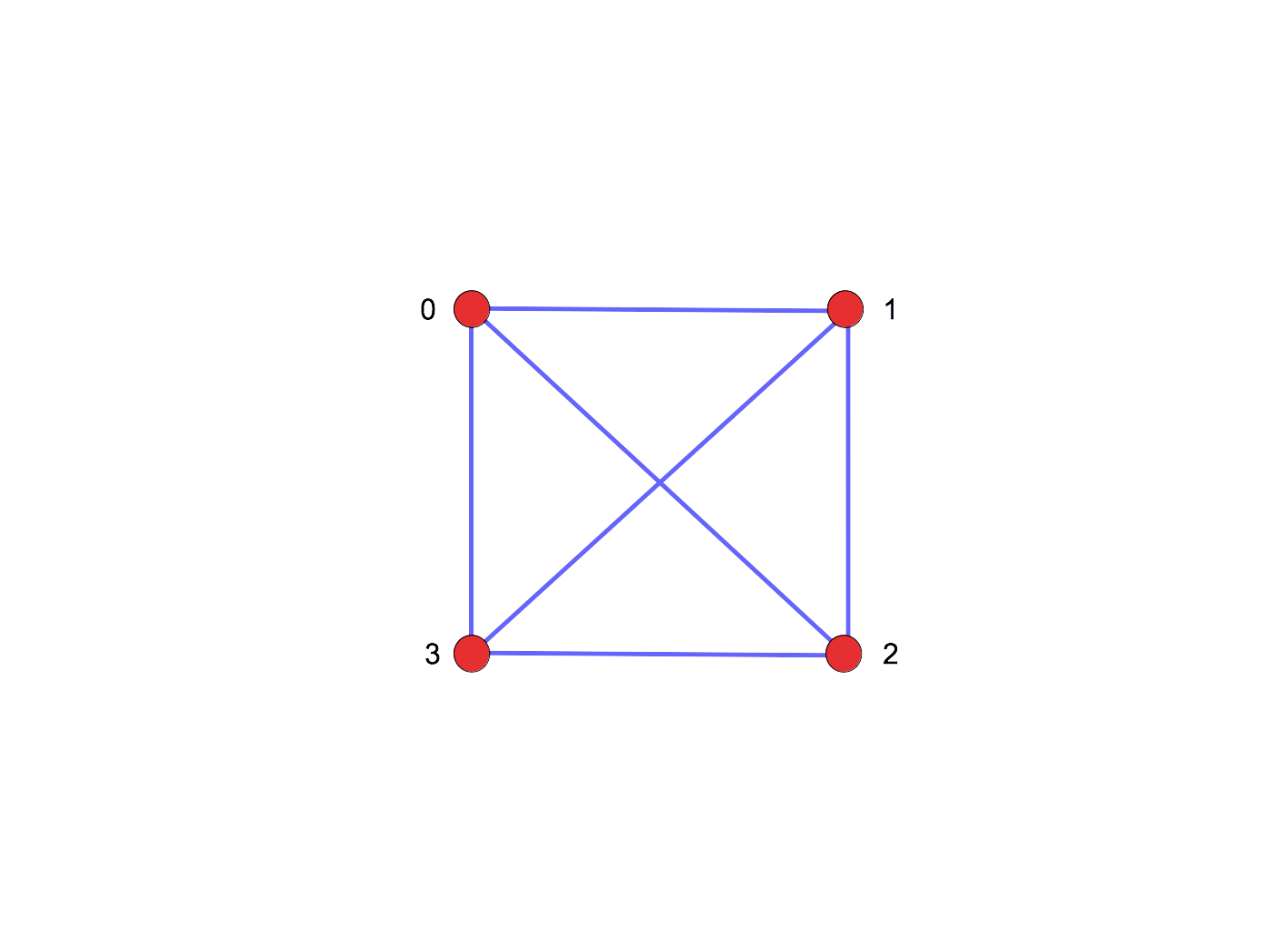} 
\end{minipage}
\vskip-1.0cm
\caption{\footnoteskip
  The left panel shows the fully connected graph $K_8$ and the right panel 
  shows the corresponding graph $K_4$. To perform a calculation to 8-bit
  accuracy requires the connectivity of $K_8$.  We take the vertex and edge 
  sets for be $K_8$ to ${\cal V}_8 = \{0, 1, 2, \cdots, 7 \}$ and ${\cal E}_8 = 
  \{ \{ 0, 1 \}, \{ 0, 2 \},  \cdots , \{ 0 , 7\} ,  \{ 1, 2 \} ,\{ 1, 3\} , \cdots,  
  \{\, 6 , 7\}\, \}$. To perform a calculation to 4-bit accuracy requires $K_4$ 
  connectivity,  and similarly, the vertex and edge sets for $K_4$ are 
  ${\cal V}_4 = \{0, 1, 2, 3 \}$ and ${\cal E}_4 = \{    \{ 0, 1\}, \{ 0, 2\}, 
  \{ 0, 3\}, \{ 1, 2\}, \{ 1, 3\}, \{ 2, 3\}    \} $.
}
\label{fig_K8_K4}
\end{figure}

For floating point division to $R$ bits of resolution, the graph ${\cal G}$ is in
fact just the fully connected graph $K_\smR$.  In terms of vertex and edge sets,
 we write  $K_\smR = ({\cal V}_\smR, {\cal E}_\smR)$, and Fig.~\ref{fig_K8_K4} 
illustrates $K_8$ and $K_4$. 
The left panel shows 
the completely connected graph $K_8$, with vertex and  edge sets 
\begin{eqnarray}
  {\cal V}_8 &=& \{0, 1, 2, \cdots, 7 \}
  \\
  {\cal E}_8 &=& 
  \{ \{ 0, 1 \}, \{ 0, 2 \},  \cdots , \{ 0 , 7\} ,  \{ 1, 2 \} , \cdots,  \{ 1, 7\} , 
  \cdots,  \{\, 6 , 7\}\, \}
  \ ,
  \nonumber \\
\end{eqnarray}
while the right panel shows the $K_4$ graph, 
\begin{eqnarray}
  {\cal V}_4 &=& \{0, 1, 2, 3 \}
  \\
  {\cal E}_4 &=& \{  \{ 0, 1\}, \{ 0, 2\}, \{ 0, 3\}, \{ 1, 2\}, \{ 1, 3\}, \{ 2, 3\}    \} 
  \ .
\end{eqnarray}
Just as 8-bits is called a {\em word}, 4-bits is called a {\em nibble}. As we will also see, 
the dynamic range of the D-Wave is most directly suitable to $K_4$, and consequently 
the connectivity of $K_4$ gives a quantum nibble. 

Let us remark about our summation conventions. Rather than summing over an edge set, 
\begin{eqnarray}
  H[Q] 
  &=&
  \sum_{r \in {\cal V}_\smR} A_r \, Q_r
  +
  \sum_{rs \in {\cal E}_\smR  }  B_{rs} \, Q_r Q_s
  \\[5pt]
  &=&
  \sum_{r=0}^{R-1} A_r\, Q_r
  +
  \sum_{r=0}^{R-1}  \sum_{s > r}^{R-1} B_{r s} \, Q_r Q_s
  \ ,
\end{eqnarray}
we find it convenient to sum over all values of $r$ and $s$ taking $B_{rs}$ to be symmetric. 
In this case, the double sum differs by a factor of two relative to summing over the edge 
set of the graph, 
\begin{eqnarray}
  H[Q] 
  &=&
  \sum_{r=0}^{R-1} A_r\, Q_r
  +
  \sum_{r=0}^{R-1}  \sum_{s = 0}^{R-1} \frac{1}{2}\, B_{r s} \, Q_r Q_s
  \ .
\end{eqnarray}
Furthermore, for $r = s$, there will be a linear contribution from the idempotency 
condition $Q_r^2 = Q_r$, so that
\begin{eqnarray}
  H[Q] 
  &=&
  \sum_{r=0}^{R-1} \left [A_r  + \frac{1}{2}\,B_{rr} \right]\, Q_r
  +
  \sum_{r=0}^{R-1}  \sum_{s \ne r , s=0}^{R-1} 
  \frac{1 }{2}\, B_{r s} \, Q_r Q_s
  \ .
\end{eqnarray}
We can write this as 
\begin{eqnarray}
  H[Q] 
  &=&
  \sum_{r=0}^{R-1}  \tilde A_r\, Q_r
  +
  \sum_{r=0}^{R-1}  \sum_{s \ne r , s=0}^{R-1} 
  \tilde B_{r s} \, Q_r Q_s
  \ .
\end{eqnarray}
\pagebreak
\section{Floating Point Division on a Quantum Annealer}

\subsection{Division as a QUBO Problem}

In this section we present an algorithm for performing floating point division on a quantum 
annealer.  Given two input parameters $m$ and $y$ to $R$-bits of resolution, the algorithm 
calculates the ratio $y/m$ to $R$ bits of resolution. The corresponding division problem 
can be represented by the linear equation
\begin{eqnarray}
  m \cdot x - y = 0
  \ ,
\label{meq_div}
\end{eqnarray}
which has the unique solution
\begin{eqnarray}
  x = y / m
  \ .
\label{sol_div}
\end{eqnarray}
Solving (\ref{meq_div}) on a quantum annealer amounts to finding an 
objective function $H(x)$ whose minimum corresponds to the solution that we 
are seeking, namely (\ref{sol_div}). Although the form of $H(x)$ is not unique, 
for this work we employ the simple real-valued quadratic function
\begin{eqnarray}
  H(x ; m, y) =  \big(m \, x - y \big)^2
  \ ,
\label{sol_H}
\end{eqnarray}
where $m$ and $y$ are continuous parmeters.  
For an ideal annealer, we do not have to concern ourselves with the numerical range 
and resolution of the parameters $m$ and $y$; however, for a real machine such as
the D-Wave, this is an important consideration. For a well-conditioned matrix,
we require that the parameters 
$m$ and $y$ possess a numerical range that spans about an order  of magnitude, from 
approximately 0.1 to 1.0. This provides about 3 to 4 bits of resolution: $1/2^0=1$, 
$1/2^1=0.5$, $1/2^2=0.25$, and $1/2^3=0.125$. The dynamic range and the 
connectivity both impact the resolution of a calculation.

To proceed, let us formulate floating point division as a {\em quadratic unconstrained 
binary optimization} (QUBO) problem. The algorithm starts by converting the 
real-valued number $x$ in (\ref{sol_H}) into an $R$-bit binary format, while the 
numbers $m$ and $y$ remain real valued parameters of the objective function.
For any number 
$\chi \in [0, 2)$, the binary representation accurate to $R$ bits of resolution can be 
expressed by $[Q_0. Q_1 Q_2 \cdots Q_{\scriptscriptstyle R-1}]_2$, where 
$Q_r\in \{0, 1\}$ is value of the $r$-th bit, and the square bracket indicates
 the binary representation.\,\footnote{\footnoteskip
Since the infinite geometric series $\sum_{r=0}^\infty 2^{-r}$ sums to 2, the finite 
series is less than 2. In binary form we have $[1.11\cdots]_2 =2$ and $[1.11 \cdots 1]_2 
< 2$. Working to resolution $R$ is like calculating the $R$-th partial sum of an infinite 
series.
}
It is more algebraically useful to express this in terms of the power series 
in $2^{-r}$,
\begin{eqnarray}
 \chi &=& \sum_{r=0}^{R-1} 2^{-r} \, Q_r
  \ .
 \label{chidef_a}
 \end{eqnarray}
In order to represent negative number, we perform the binary offset
\begin{eqnarray}
  x &=& 2 \chi - 1
  \ ,
  \label{xdef}
\end{eqnarray}
where $x \in [-1, 3)$. The objective function now takes the form
\begin{eqnarray}
  H(\chi)
  &=&
  4 m^2 \chi^2 - 4 m (m + y) \chi + (m + y)^2
 \ .
\label{H_chi_a}
\end{eqnarray}
The constant term $(m+y)^2$ can be dropped when finding the minimum
of (\ref{H_chi_a}), but we choose to keep it for completeness.
Equation~(\ref{chidef_a}) provides a change of variables $\chi=\chi[Q]$ 
(where $Q$ is the collection of the $Q_r$), and this allows  us to express 
(\ref{sol_H}) in the form
\begin{eqnarray}
  H[Q] 
  &=&
  \sum_{r=0}^{R-1} A_r \, Q_r
 + \sum_{r=0}^{R-1} \sum_{s \ne r, s = 0}^{R-1} B_{rs} \,  Q_r  Q_s
 \ .
\label{hsum_div_logical}
\end{eqnarray}
In the notation of graph theory, we would write
\begin{eqnarray}
  H[Q]
  &=&
  \sum_{r \in {\cal V}_\smR} A_r\, Q_r
  +
  \sum_{rs \in {\cal E}_\smR} \frac{1}{2}\, B_{r s} \,  Q_r  Q_s
  \ ,
  \label{QUBO_sec2}
\end{eqnarray}
where 
${\cal V}_\smR = \{0, 1, 2, \cdots,  R-1 \}$ is the vertex set, and ${\cal E}_\smR$ 
is the edge set. We often employ an abuse of notation and write $rs \in {\cal E}_\smR$ 
to mean $\{r,s\} \in {\cal E}_\smR$. We should also use the notation $B_{ \{r, s\} }$, 
but instead we write $B_{rs}$. Since the
order of the various elements of a set are immaterial, we require $B_{rs}$ to be symmetric 
in $r$ and $s$.  Rather than summing over the edge sets $r s \in {\cal E}_\smR$, we employ 
the double sum $\sum_{r \ne s}$, which introduces a relative factor of two in the convention 
for the strengths $B_{r s}$. The goal of this section is to find $A_r$ and $B_{rs}$ in terms of 
$m$ and $y$.

Note that we can generalize the simple binary offset (\ref{chidef_a}) if we scale and shift 
$\chi \in [0,2)$ by
\begin{eqnarray}
  x &=& c \chi - d
  \ ,
\label{xdefb_two}
\end{eqnarray}
so that $x \in [-d, 2 c - d)$. When $d > 0$ and $c > d/2$, the domain of $x$ 
always contains a positive and negative region, and the precise values for $d$ 
and $c$ can be chosen based on the specifics of the problem.
For Eq.~(\ref{xdefb_two}), the objective function takes the form
\begin{eqnarray}
  H(\chi)
  &=&
  4 m^2 c^2\, \chi^2 - 4 m c \, (m + y) \chi + (m d + y)^2
 \ .
\label{H_chi_xdefB}
\end{eqnarray}
For simplicity of notation, this paper employs the simple binar offset (\ref{xdef}),
although our Python interface to the D-Wave quantum annealer employs the generalized 
form (\ref{H_chi_xdefB}).

Equation~(\ref{chidef_a}) allows us to express the quadratic term in $\chi$ as
\begin{eqnarray}
  \chi^2 
  =
  \sum_{r=0}^{R-1} \sum_{s=0}^{R-1} 2^{-r-s} Q_r Q_s
  =
  \sum_{r=0}^{R-1} \sum_{s \ne r,  s=0}^{R-1} 2^{-r-s } Q_r Q_s
  +
  \sum_{r=0}^{R-1} 2^{-2 r } Q_r
  \label{chi_squared}
  \ ,
\end{eqnarray}
where we have used the idempotency condition $Q_r^2 = Q_r$ along the diagonal
in the last term of (\ref{chi_squared}). Substituting the forms (\ref{chidef_a}) and
(\ref{chi_squared}) into (\ref{H_chi_a}) provides the Hamiltonian 
\begin{eqnarray}
  H[Q]
  &=&
  \sum_{r=0}^{R-1} 4 m\, 2^{-r}\Big[m\, 2^{-r} - (y + m)\Big] Q_r
    +
  \sum_{r=0}^R\sum_{s \ne r, s =0}^{R-1} 4 m^2 \, 2^{-r-s}\, Q_r Q_s
 \ ,
\label{Hq}
\end{eqnarray}
and the Ising coefficients in (\ref{hsum_div_logical}) can be read off:
\begin{eqnarray}
  A_r &=& 4 m\, 2^{-r}\Big[ m\, 2^{-r} - (y + m) \Big] 
  \\[5pt]
 B_{rs} &=& 4 m^2 \, 2^{-r-s}  ~~~ r \ne s
 \ .
\label{ABq}
\end{eqnarray}
Because of the double sum over $r$ and $s$ in the objective function in (\ref{Hq}),
the algorithm requires a graph of connectivity $K_\smR$. The special cases of $K_8$ 
and $K_4$ have been illustrated in Fig.~\ref{fig_K8_K4}.  To obtain higher accuracy 
than the $K_\smR$ graph allows, we can iterate this procedure in the following 
manner. Suppose we start with $y_0 = y$, and we are given a value $y_{n-1}$ 
with $n > 1$, then we advance the iteration to $y_{n}$ in the following manner,
\begin{eqnarray}
 {\rm solve}~~ m x_n &=& y_{n-1}  ~~{\rm for}~ x_n ~{\rm to}~ R ~{\rm bits}
  \\
  {\rm calculate~the~error}~~  y_n &=& y_{n-1} - m x_n 
  \ .
\label{iterate_y}
\end{eqnarray}
Now that we have the value $y_n$, we can repeat the process to find $y_{n+1}$,
and we can stop the iterative procedure when the desired level of accuracy has 
been achieved. 

\subsection{Embedding $\bm{K_{\scriptscriptstyle R}}$ onto the D-Wave Chimera Architecture}
\label{sec_embedding}

The D-Wave Chimera chip consists of coupled bilayers of micro rf-SQUIDs overlaid in such a way 
that, while relatively easy to fabricate, results in a fairly limited set of physical connections between 
the qubits. However, by {\em chaining} together well chosen qubits in a positively correlated manner, 
this limitation can largely be overcome. The process of chaining requires that we (i) embed the 
logical graph onto the physical graph of the chip (for example $K_4$ onto $C_8$) and that we 
(ii) assign weights and strengths to the physical graph embedding in such as a way as to preserve 
the ground state of the logical system. These steps are called graph embedding and Hamiltonian 
embedding, respectively. 

\begin{figure}[h!]
\begin{minipage}[c]{0.4\linewidth}
\includegraphics[scale=0.40]{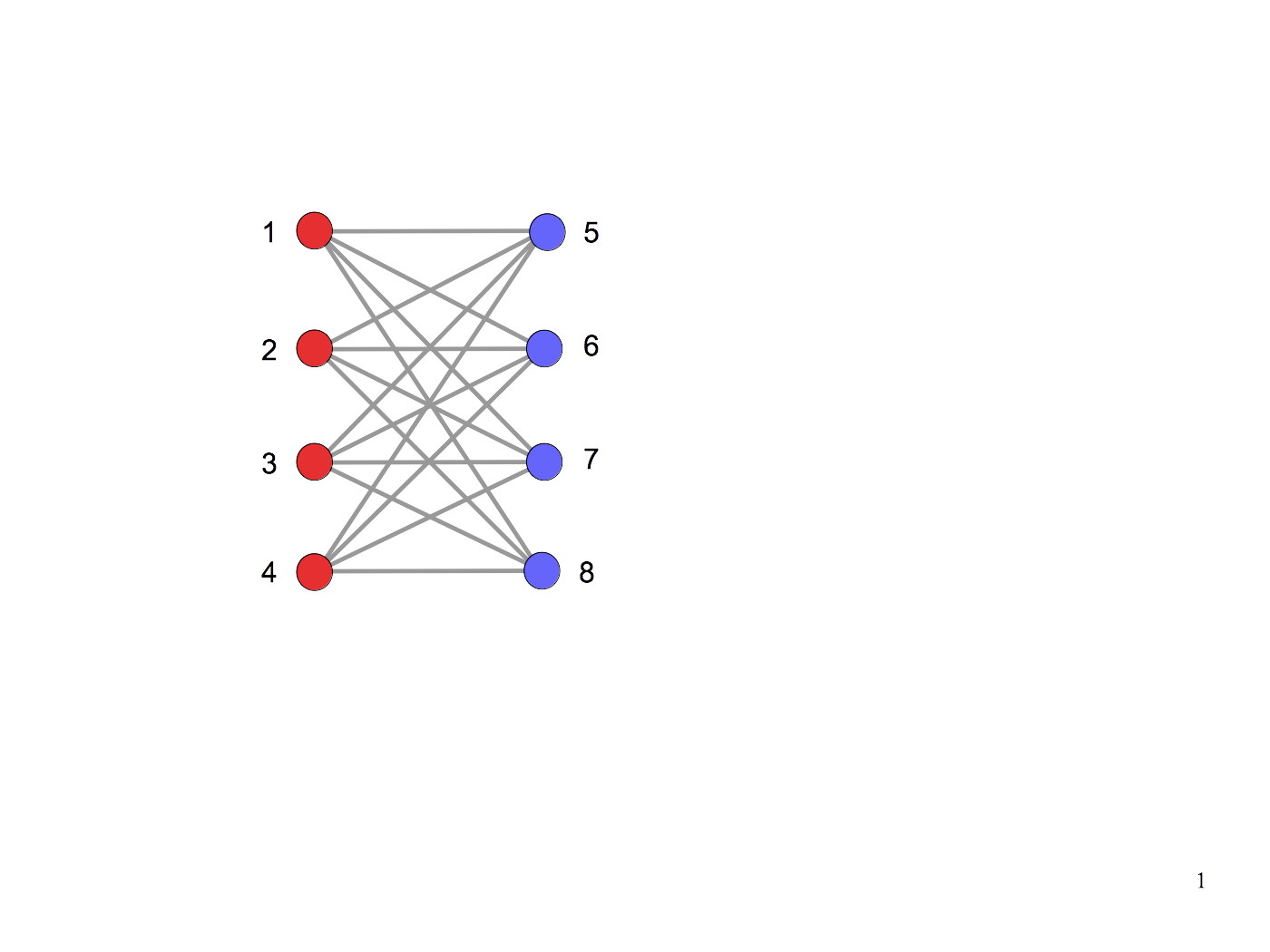}
\end{minipage}
\hfill
\begin{minipage}[c]{0.4\linewidth}
\includegraphics[scale=0.35]{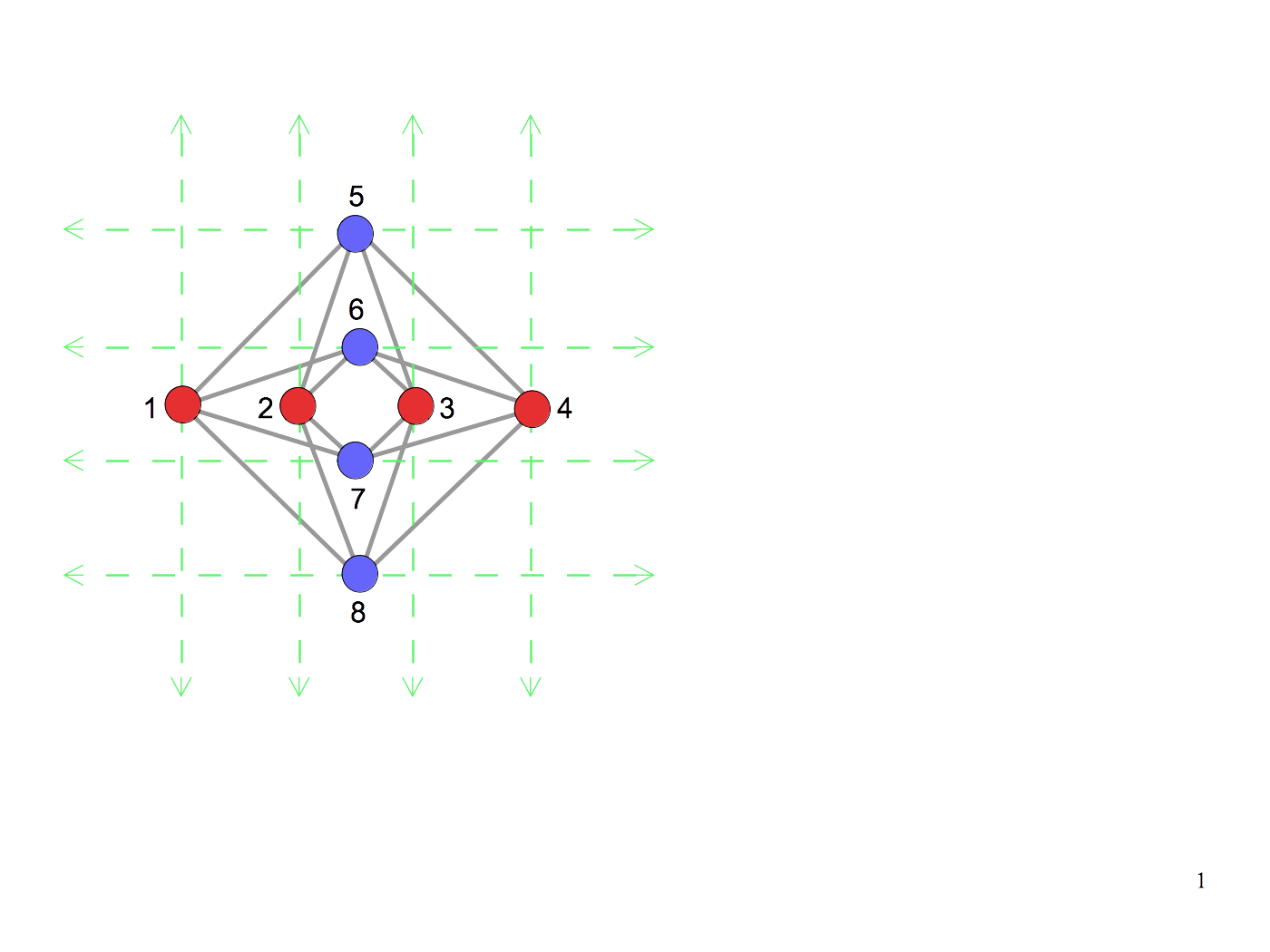}
\end{minipage}
\vskip-2.0cm
\caption{\footnoteskip
The left panel illustrates the bipartate graph $C_8$ in {\em column} format, 
while the right panel illustrates the corresponding graph in {\em cross} format,
often called a Chimera graph. The gray lines represent direct connections 
between qubits. The cross format is useful since it minimizes the number 
intersecting connections. The use of red and blue dots emphasize the bipartate 
nature of $C_8$, as every red dot is connected to every blue dot, while none 
of the red and blue dots are connected to one another. The vertex set of
$C_8$ is taken to be ${\cal V}_8 = \{1, 2, \cdots, 8 \}$ and edge set is 
${\cal B}_8 = \{ \{1, 5\} , \{1, 6\} , \{1, 7\} , \{1, 8\} , \{2, 5\} ,\{2, 6\} 
\cdots \{ 7, 8\} \}$. 
}
\label{fig_chimera_topology}
\end{figure}
\begin{figure}[h!]
\begin{minipage}[c]{0.4\linewidth}
\includegraphics[scale=0.30]{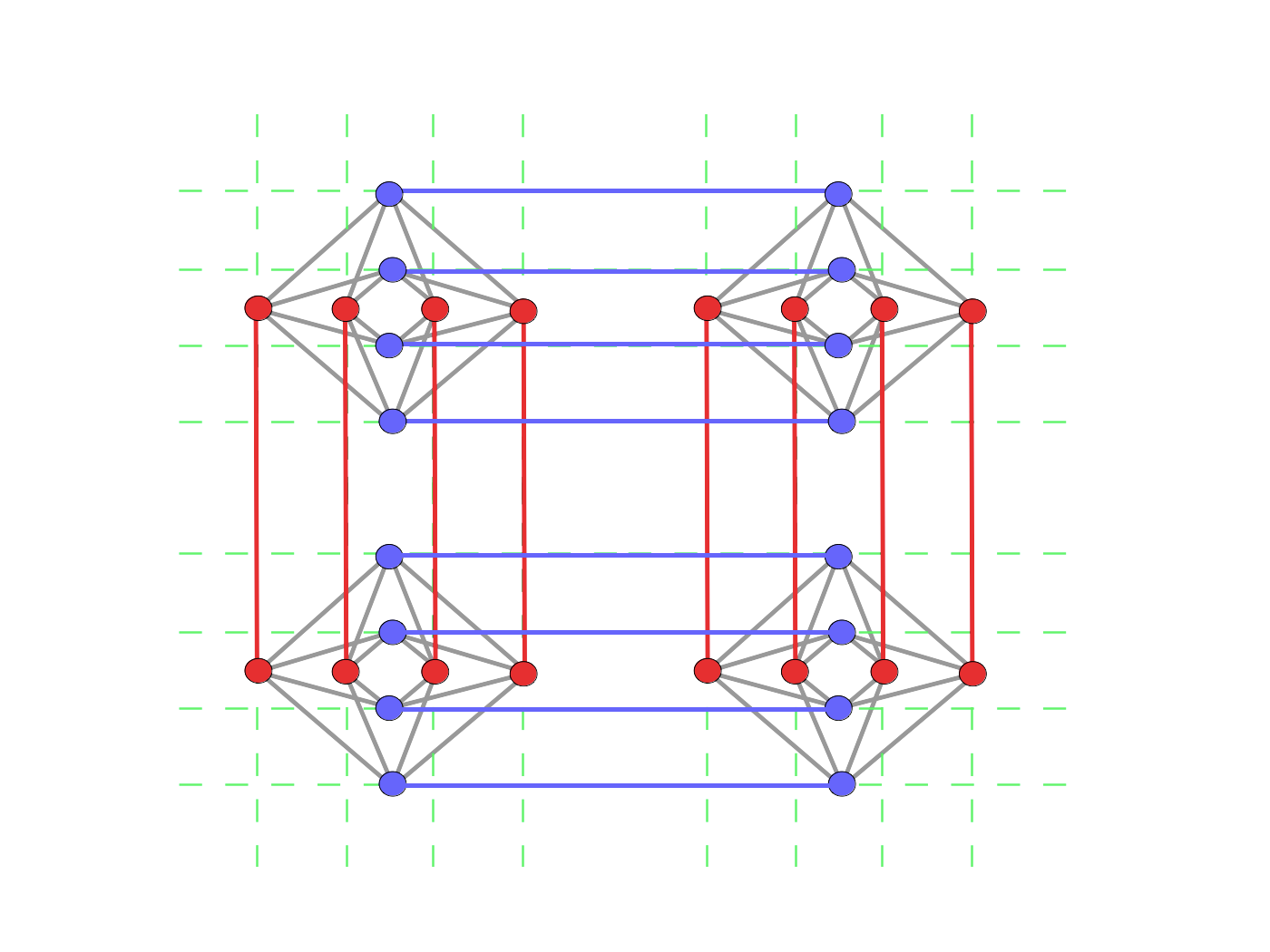}
\end{minipage}
\hfill
\begin{minipage}[c]{0.4\linewidth}
\includegraphics[scale=0.28]{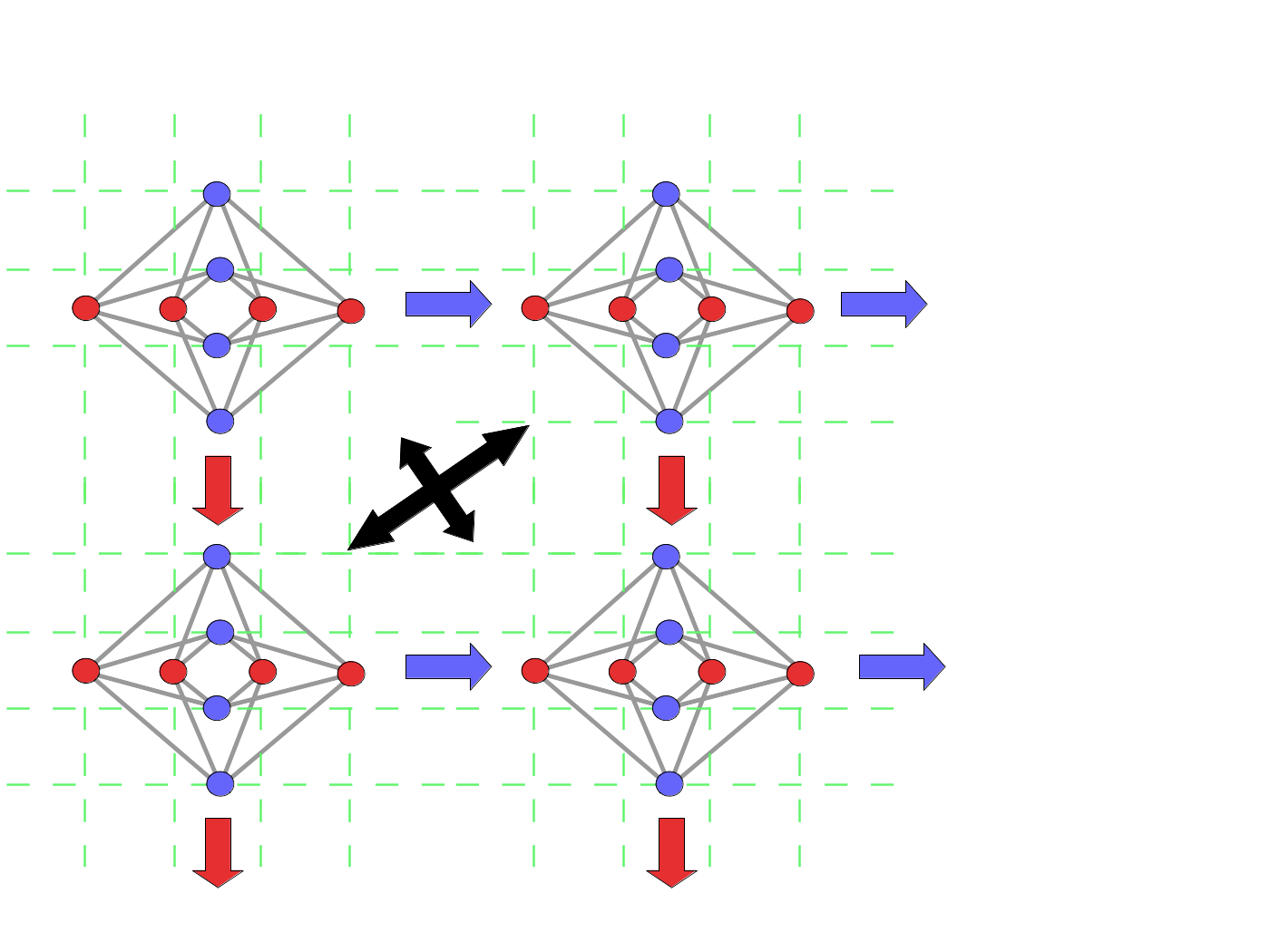}
\end{minipage}
\caption{\footnoteskip
The the left panel shows the connectivity between four $C_8$ bipartate 
Chimera zones, and the right panel illustrates how multiple $C_8$
graphs are stitched together along the vertical and horizontal
directions to provide thousands of possible qubits. A limitation of
this connectivity strategy is that red and blue zones cannot
communicate directly with one another, as indicated by the black
crossed arrows. The purpose of {\em chaining} is to allow
communication between the read and blue qubits.  
}
\label{fig_bipartate8_connect}
\end{figure}
Let us explore the connectivity of the D-Wave Chimera chip in more detail. The D-Wave 
architecture employs the $C_8$ bipartate Chimera graph as its most basic unit of connectivity.  
This {\em unit cell } is illustrated in Fig.~\ref{fig_chimera_topology}, and consists of 8 qubits 
connected in a $4 \times 4$ bipartate manner. The left panel of the figure uses a {\em column} 
format in laying out the qubits, and the right panel illustrates the corresponding qubits in a 
{\em cross} format, where the gray lines represent the direct connections between the qubits. 
The cross format is useful since it minimizes the number intersecting connections. The complete 
two dimensional chip is produced by replicating  $C_8$ along the vertical and horizontal 
directions, as illustrated in Fig.~\ref{fig_bipartate8_connect}, thereby providing a chip with 
thousands of qubits. The connections between qubits are limited in two ways: (i) by the 
connectivity of the basic unit cell $C_8$ and (ii) by the connectivity between the unit cells 
across the chip. The bipartate graph $C_8 = ( {\cal V}_8, {\cal B}_8 )$ is formally defined by
the vertex set ${\cal V}_8=\{1, 2, \cdots, 8 \}$, and the edge set
\begin{eqnarray}
  {\cal B}_8 
  &=& 
  \big\{ \{1, 5\} , \{1, 6\} , \{1, 7\} , \{1, 8\} , 
  \{2, 5\} ,\{2, 6\} , \{2, 7\}  , \{2, 8\} ,
  \nonumber
  \\
  && 
  ~~
  \{3, 5\} ,\{3, 6\} , \{3, 7\}  , \{3, 8\} ,
  \{4, 5\} ,\{4, 6\} , \{4, 7\}  , \{4, 8\} 
  \big\}. 
  \label{eq_bipartate_eight}
\end{eqnarray}
The set ${\cal B}_8$  represents the connections between a given red qubit and 
the corresponding blue qubits in the Figures. The red and blue dots illustrate the 
bipartate nature of $C_8$, as every red dot is connected to every blue dot, while 
none of the blue and red dots are connected to one another. 

\begin{figure}[h!]
\begin{minipage}[c]{0.4\linewidth}
\includegraphics[scale=0.40]{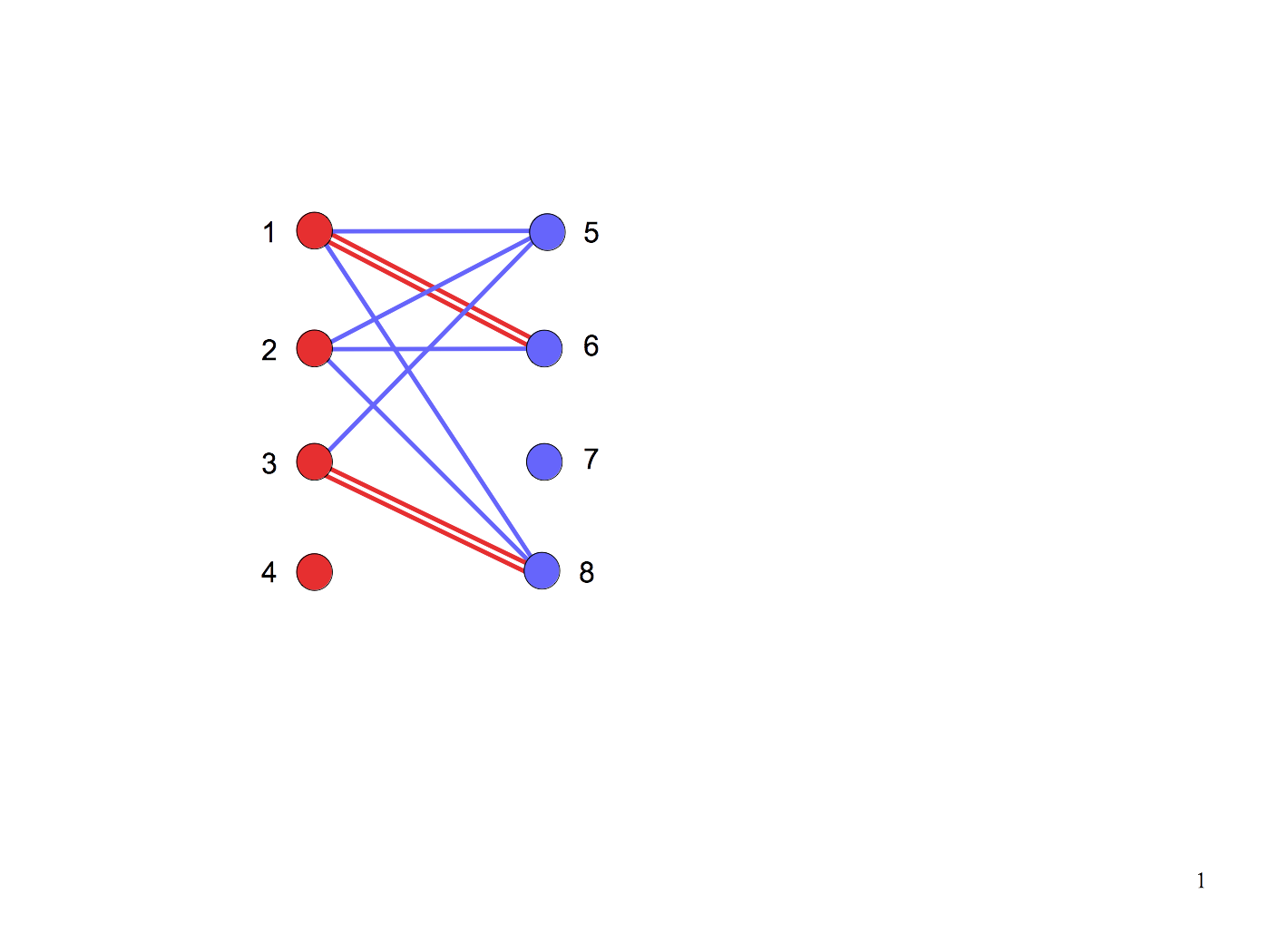}
\end{minipage}
\hfill
\begin{minipage}[c]{0.4\linewidth}
\includegraphics[scale=0.35]{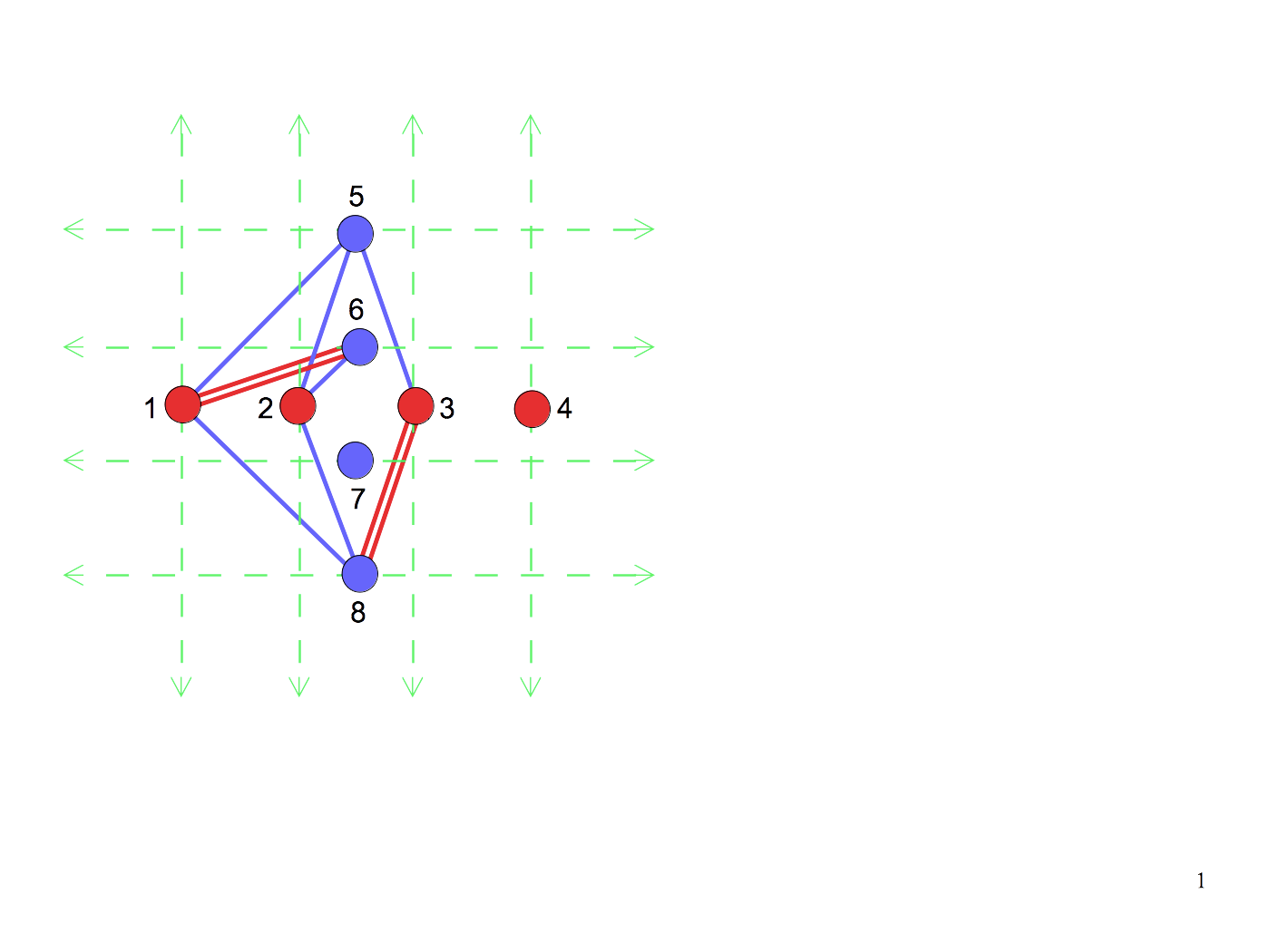}
\end{minipage}
\vskip-2.0cm
\caption{\footnoteskip
The $K_4$ embedding onto $C_8$ used in our implementation 
of 4-bit of division on the D-Wave. The blue lines represent normal connections 
between qubits, while the red double-lines represent chained qubits, that is 
to say, qubits that are strictly correlated (and can thereby represent a  single 
logical qubit at a higher level of abstraction). The qubits 1-6 are chained
together, as are the qubits 3-8.
}
\label{fig_K4_embedding_1}
\end{figure}

We will denote the {\em physical} qubits on the D-Wave chip by $q_\ell$. For 
the D-Wave 2000Q there is a maximum of 2048 qubits, while the D-Wave 2X 
has 1152 qubits. For the example calculation in this text, we only use 10 to 
50 qubits. The physical Hamiltonian or objective function takes the form
\begin{eqnarray}
  H[q] = \sum_\ell a_\ell \, q_\ell + \sum_{\ell \ne m} 2 b_{\ell m} \, q_\ell q_m
  \ ,
\end{eqnarray}
where we have introduced a factor of 2 in the strength  to account for the 
symmetric summation over $r$ and $s$. We will call the qubits $Q_r$
of the previous section the {\em logical  qubits}. To write a program for the
D-Wave means finding an embedding of the logical problem onto
the physical collection of qubits $q_\ell$. If the connectivity of the Chimera 
graphs were large enough, then the logical qubits would coincide exactly with 
the physical qubits. However, since the graph $C_8$ possesses less connectivity 
than $K_4$,  we must resort to chaining on the D-Wave, even for
4-bit resolution. Figure~\ref{fig_K4_embedding_1} illustrates the $K_4$ embedding 
used by our algorithm, where, as before, the left panel illustrates the bipartate 
graph in column format, and the right panel illustrates the corresponding graph 
in cross format. 
\begin{figure}[h!]
\includegraphics[scale=0.40]{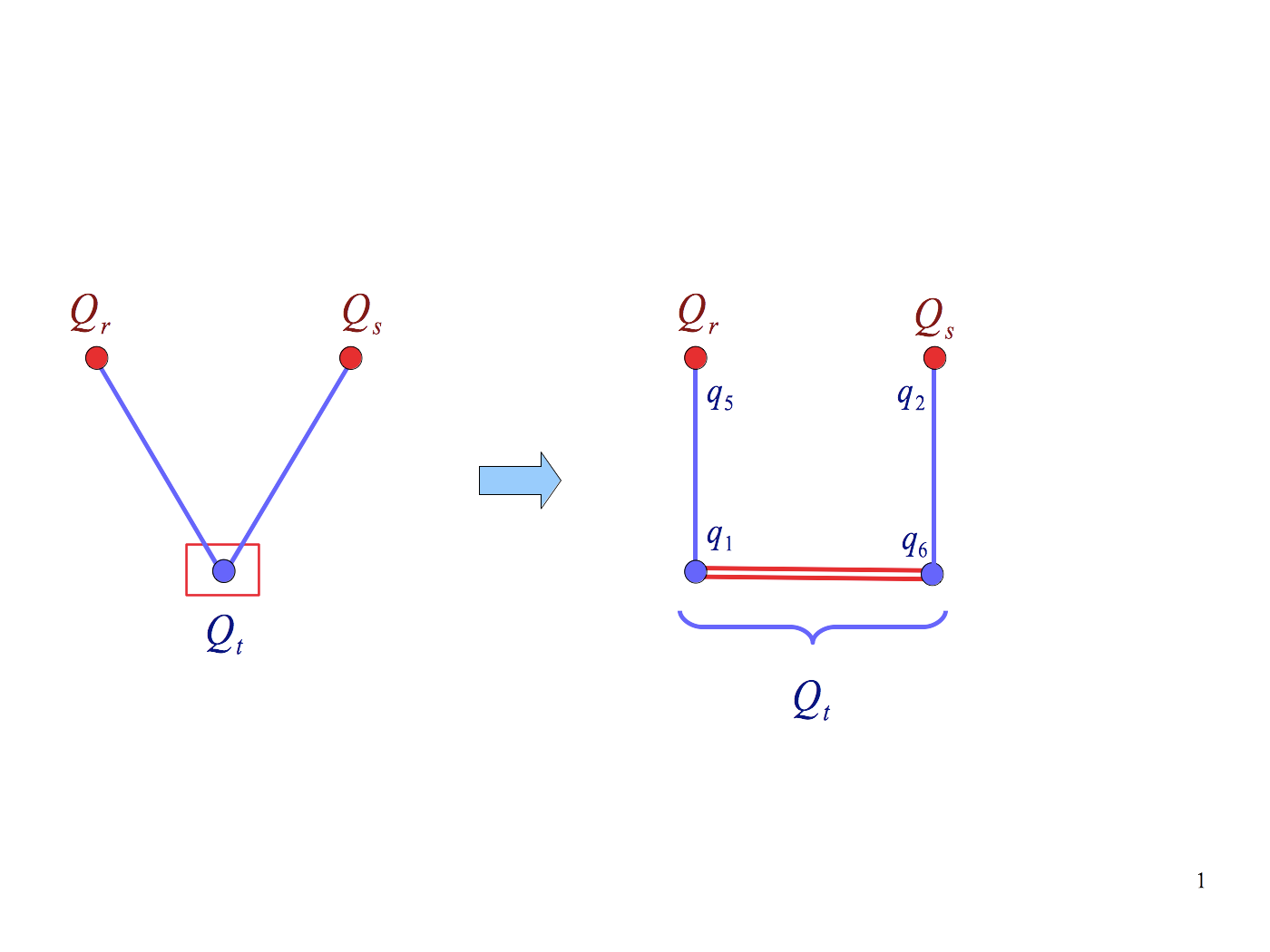} 
\vskip-2.0cm
\caption{\footnoteskip  
The left panel shows three logical qubits $Q_r$, $Q_s$, $Q_t$ with connectivity 
between $r$-$t$ and $t$-$s$. The box surrounding qubit $t$ means that it
will be modeled by a linear chain of physical qubits, as illustrated in the right
panel. The labeling is taken from Fig.~\ref{fig_K4_embedding_1} for qubits
5-1-6-2, where $Q_r$ is mapped to $q_5$, $Q_s$ is mapped to $q_2$, and
$Q_t$ is split between $q_1$ nd $q_6$. Qubits $q_1$ and $q_6$ are chained 
together to simulate the single logical qubit $Q_t$, while qubits $Q_r$ and 
$Q_s$ map directly onto physical qubits $q_5$ and $q_2$.
}
\label{fig_linear_chain_a}
\end{figure}

In Fig.~\ref{fig_K4_embedding_1} we have labeled the physical qubits by 
$\ell=1, 2, 3 \cdots 8$, and we wish to map the logical problem involving 
$Q_r \, Q_s \, Q_t$  onto the four physical qubits $q_5 \,q_1 \, q_6 \, q_2$. 
The embedding requires that we {\em chain}  together  the two qubits 1-6 
and 3-8, respectively. We may omit qubits $4$ and $7$ entirely. 
As illustrated in Fig.~\ref{fig_linear_chain_a}, the physical qubits $q_1$ and $q_6$ 
are {\em chained} together to simulate a single logical qubit $Q_t$, while qubits 
$q_5$ and $q_2$ are mapped directly to the logical qubits $Q_r$ and $Q_s$, 
respectively. Qubit $q_5$ is assigned the weight $a_5 = A_r$ and the coupling 
between $q_5$ and $q_1$ is assigned the value $b_{51}=B_{rt}$. Similarly for 
qubit $q_2$, the vertex is assigned weight $a_2 =  A_s$, and strength between 
$q_2$ and $q_6$ is $b_{26}=B_{st}$. We must now distribute the logical qubit 
$Q_t$ between $q_1$ and $q_6$ by assigning the values $a_1$, $a_6$ and 
$b_{16}$. We distribute the weight $A_t$ uniformly between qubits $q_1$ and 
$q_2$, giving $a_1 = A_t/2$ and $a_6 = A_t/2$. We must now choose $b_{16}$.
To preserve the energy spectrum, we must shift the values of the weights $a_1$
and $a_6$. We can do this by adding a counter-term Hamiltonian
\begin{eqnarray}
  H^\smCT= a \, q_1 + a\, q_6 + 2 b_{16} \, q_1 q_6
  \ 
\end{eqnarray}
to the physical Hamiltonian. 
\begin{figure}[h!]
\begin{minipage}[c]{0.4\linewidth}
\includegraphics[scale=0.25]{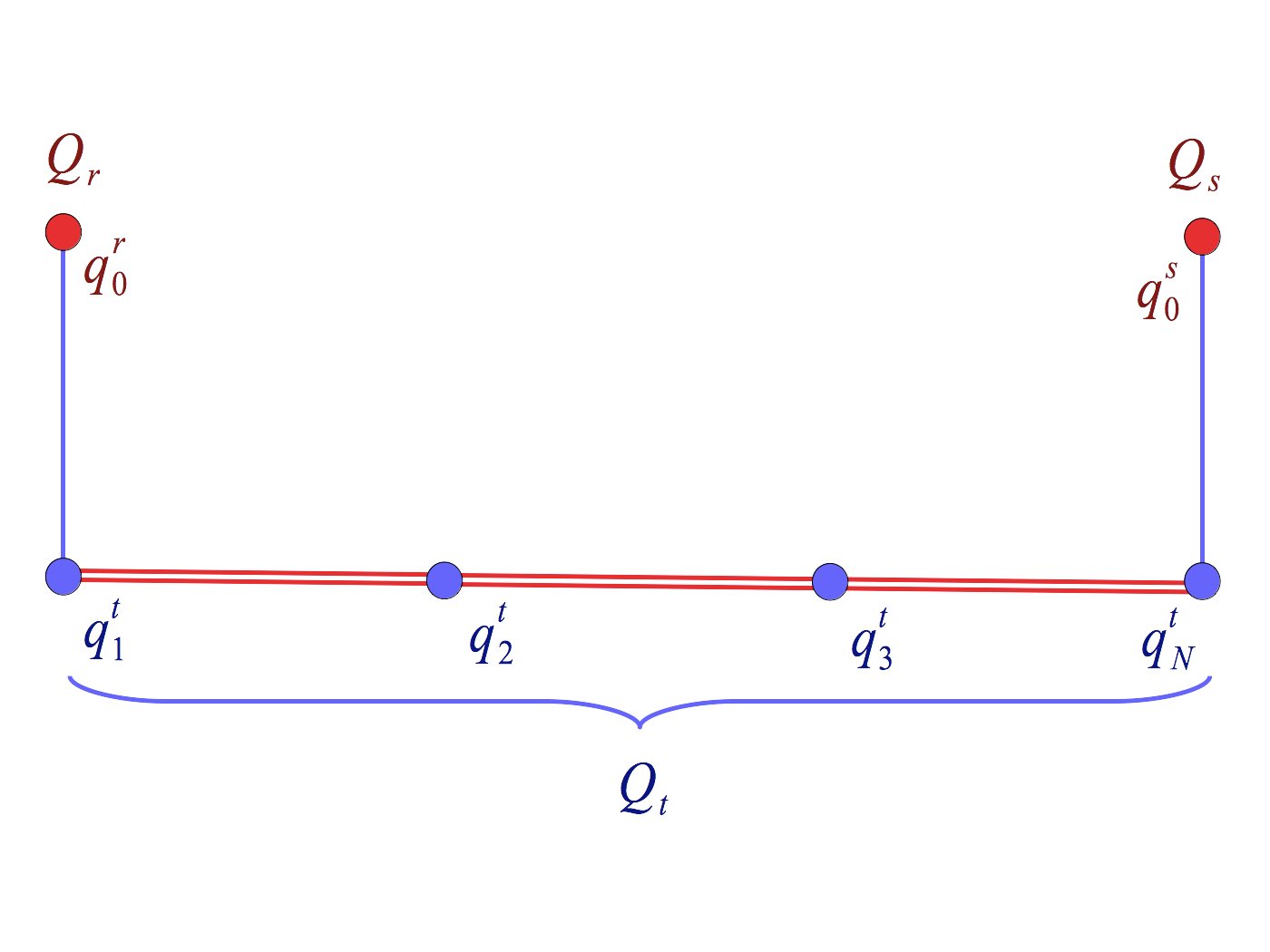}
\end{minipage}
\hfill
\begin{minipage}[c]{0.48\linewidth}
\includegraphics[scale=0.30]{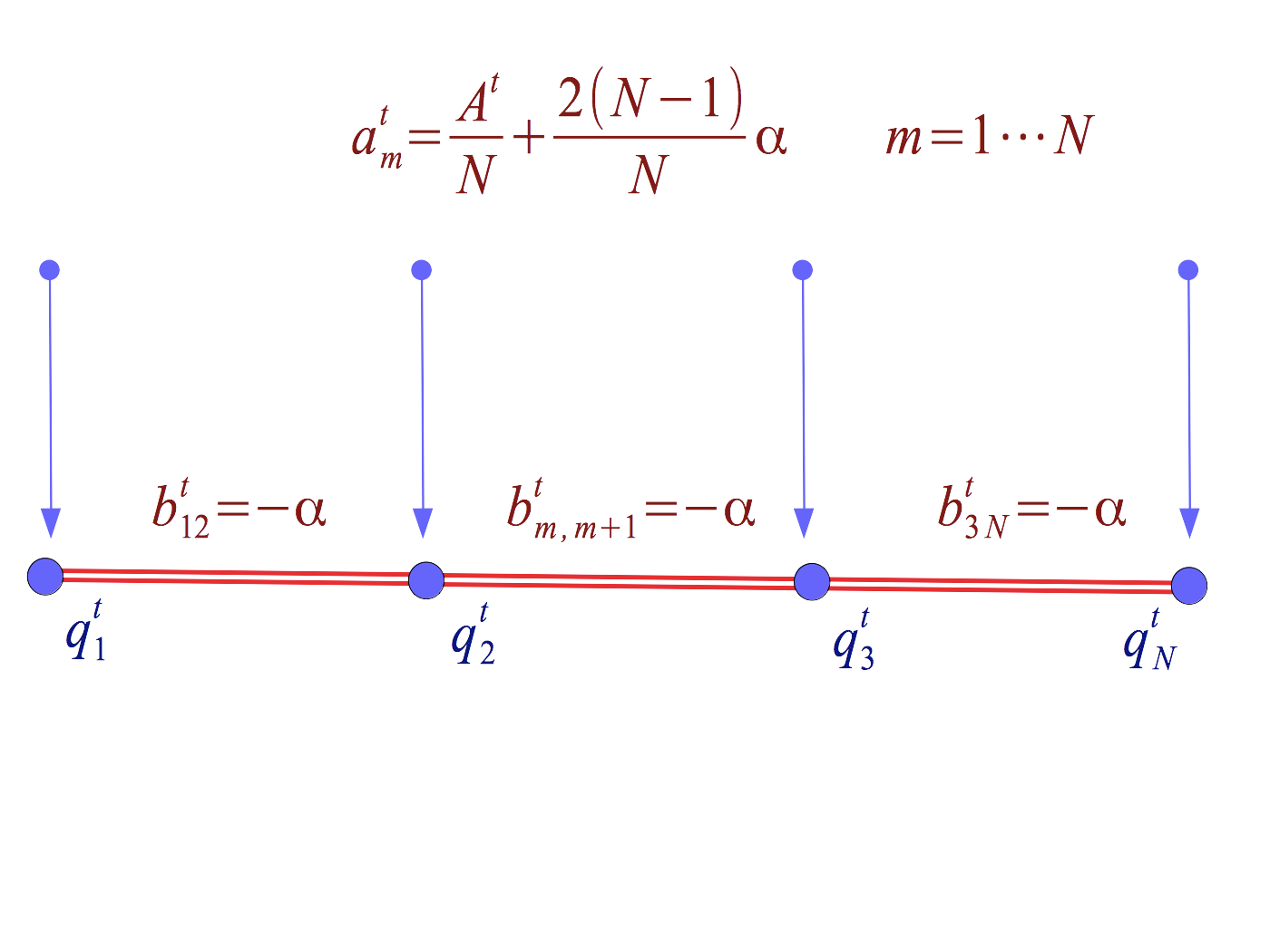}
\end{minipage}
\vskip-1.0cm
\caption{\footnoteskip
Generalization of Fig.~\ref{fig_linear_chain_a} to a chain of $N$ linear 
qubits. 
The right panel illustrates the chain coupling parmeters used to create strict 
correlations of the physical qubits within the chain. 
}
\label{fig_linear_chain_b}
\end{figure}
The double lines in  Figs.~\ref{fig_K4_embedding_1} and \ref{fig_linear_chain_a} indicate 
that two qubits are chained together. 
This means that the qubits are strictly correlated, {\em i.e.} when $q_1$ is up then 
$q_6$ is up, and when $q_1$ is down then $q_6$ is down. This is accomplished by 
choosing the coupling strength $b_{16}$ to favor a strict correlation; however, 
to preserve the ground state energy, this also requires shifting the weights for $q_1$
and $q_6$. 
For $q_1=q_6=0$ we have $H^\smCT=0$. We wish to preserve this condition when $q_1=
q_6=1$, which means $2 a + 2 b = 0$. Furthermore, the state $q_1=1$ and $q_6=0$ must 
have positive energy, which means $a > 0$. Similarly for $q_1=0$ and $q_6=1$. We therefore 
choose $a_1 = a_6 = \alpha >0$ and $b_{16} = -\alpha$, where $\alpha$ is an arbitrary
parameter. This is illustrated in  Table.~\ref{table_HCT_spectrum_2}.
\begin{table}[t!]
\caption{\footnoteskip 
  For two qubits the counter-term Hamiltonian is $H^{\smCT}(q_1, q_6) = a\,  q_1 + a\,  
  q_6 + 2 b\,  q_1 q_6$. The lowest energy state is preserved for $b = - \alpha$  and 
  $a = \alpha$ where $\alpha > 0$. We will split the weight $A_t$ uniformly across the 
  $N$ chained physical qubits, thereby giving a contribution to the physical Hamiltonian  
  $H_{16}^t  = A_t / 2 + \alpha\,  q_1 + \alpha\,  q_6 - 2 \alpha\,  q_1 q_6$. The energy
  spectrum ensures that the two qubits are strictly correlated. 
}
\vskip0.3cm
\begin{tabular}{|c|c||l|} \hline
   ~$q_1$~    & ~$q_6$~    & ~$H^{\smCT}$~  \\\hline
   0 & 0 & \,0 \\[-8pt]
   0 & 1 & \,$\alpha$ \, \\[-8pt]
   1 & 0 & \,$\alpha$ \,  \\[-8pt]
   1 & 1 & \,0  \\ \hline
\end{tabular} 
\label{table_HCT_spectrum_2}
\end{table}
A more complicated case is the linear chain of $N$ qubits as shown in 
Fig.~\ref{fig_linear_chain_b}. The counter-term Hamiltonian is taken to be
\begin{eqnarray}
  H^\smCT = \sum_{m=1}^N a_m^t \, q^t_m + \sum_{m=1}^{N-1} b_{m, m+1}^t \,
   q_m^t q_{m+1}^t
  \ .
\label{eq_ct_N}
\end{eqnarray}
Note that $H^\smCT$ vanishes when $q_m=0$ for all $m=1 \cdots N$. And conversely, we
must arrange the counter-term to vanish when $q_m=1$. The simplest choice is to take 
all weights to be the same and all couplings to be identical. Then, to preserve the ground
state when the $q_r=1$, we impose
\begin{eqnarray}
  a^t_r = \frac{A_t}{N} + \frac{2(N-1)}{N} \, \alpha
  \\[8pt]
  b^t_{r ,r+1} = - \alpha
\label{eq_name}
\end{eqnarray}
with $\alpha > 0$ and $r=1 \cdots  N$. The first term in $a^t_r$ distributes the 
weight $A_t$ uniformly across all $N$ nodes in the chain. The second set  
of terms $b^t_{r, r+1}$ ensures that the qubits of the chain are strictly correlated. 
The counter-term energy is positive and is therefore selected against when the linearly 
chained qubits are not correlated. 
Table~\ref{table_HCT_spectrum_3} illustrates the spectrum of the counter-term 
Hamiltonian for three qubits.  We often need to choose large
values of $\alpha$, of order 20 or more, to sufficiently separate the 
states. The uniform spectrum of 4 states with $H^{\smCT}=a$ in 
Table~\ref{table_HCT_spectrum_3} arises from a permutation 
symmetry in $q_1$, $q_2$, $q_3$.
\begin{table}[h!]
\caption{\footnoteskip 
  For a three qubit chain the counter-term Hamiltonian is 
  $H^{\smCT}(q_1, q_2, q_3) = a\,  q_1 + a\,  q_2 + a\,  q_3 + 
  2 b\,  q_1 q_2 + 2 b\,  q_2 q_3$, where $a=4 \alpha/3$ and
  $b=-\alpha$. The degeneracy in energy of value $a$ arises 
  from a permutation symmetry in $q_1 \to q_2 \to q_3$ that 
  preserves the form of the counter-term Hamiltonian. 
}
\vskip0.3cm
\begin{tabular}{|c|c|c||c|l|} \hline
  ~$q_1$~  &  ~$q_2$~ & ~$q_3$~ & ~$H^\smCT$ & ~ \\\hline
   0 & 0 & 0 & 0  & ~0~\\[-8pt] 
   0 & 0 & 1 & ~$4\alpha/3$~ & ~$a$~ \\[-8pt] 
   0 & 1 & 0 & ~$4\alpha/3$~ & ~$a$~\\[-8pt] 
   0 & 1 & 1 & ~$2\alpha/3$~ & ~$a/2$~ \\[-8pt] 
   1 & 0 & 0 & ~$4\alpha/3$~ & ~$a$~ \\[-8pt] 
   1 & 0 & 1 & ~$8\alpha/3$~ & ~$2a$~ \\[-8pt] 
   1 & 1 & 0 & ~$2\alpha/3$~ & ~$a$~ \\[-8pt] 
   1 & 1 & 0 & 0 & ~0~ \\\hline
\end{tabular} 
\label{table_HCT_spectrum_3}
\end{table}

\begin{figure}[h!]
\includegraphics[scale=0.35]{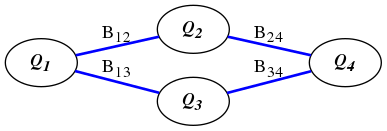} 
\caption{\footnoteskip  
Four logical qubits $Q_1$, $Q_2$, $Q_3$, $Q_4$ in a circular loop
with connection strengths $B_{12}$, $B_{2 4}$ $B_{4 3}$ and $B_{3 1}$. 
}
\label{fig_logical_qubits}
\end{figure}

\begin{figure}[h!]
\includegraphics[scale=0.35]{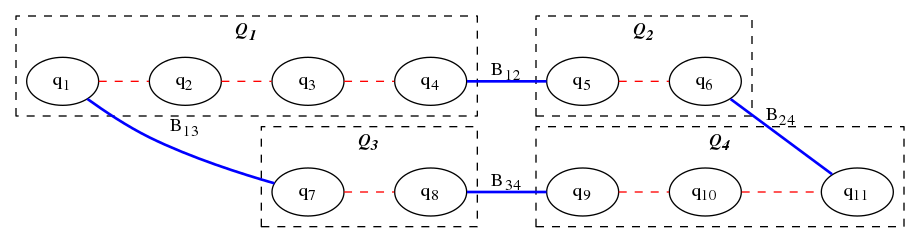} 
\caption{\footnoteskip  
A possible mapping of the logical qubits in Fig.~\ref{fig_logical_qubits} onto
the physical device. Each logical qubit is modeled by a linear chain of strictly 
correlated qubits.
}
\label{fig_embedded_qubits}
\end{figure}

To review, note that a linear counter-term is represented in Fig.~\ref{fig_linear_chain_b}. 
We add a counter-term to break the logical qubits into a chain of physical qubits that 
preserve the ground state.
Let us consider the conditions that we place on the Hamiltonian  to
ensure strict correlation between the chained qubits. We adjust the values of $A_r$
and $B_{rs}$ to ensure that spin alignment is energetically favorable. By slaving several 
qubits together, we can overcome the limitations of the Chimera connectivity. 
As a more complex  example, consider the four logical qubits of 
Fig.~\ref{fig_logical_qubits} connected in a circular chains by strengths $B_{12}$, 
$B_{2 4}$, $B_{4 3}$, and $B_{3 1}$.  Suppose the weights are $A_1$, $A_2$
$A_3$ and $A_4$. Figure~\ref{fig_embedded_qubits} provides an example 
in which each logical qubit is chained in a linear fashion to the physical qubits. 
%

%
%
%
%
%
%
%
\pagebreak
\clearpage
\section{Matrix Inversion as a QUBO Problem}

In this section we present an algorithm for solving a system of linear equations on a 
quantum annealer. To precisely define the mathematical problem, let $\boldsymbol{M}$
be a nonsingular $N \times N$ real matrix, and  let $\bvec{Y}$ be a real $N$
dimensional vector; we then wish to solve the linear equation 
\begin{eqnarray}
  \boldsymbol{M} \cdot \bvec{x} =  \bvec{Y}
  \ .
\label{meq}
\end{eqnarray}
The linearity of the system means that there is a unique solution,
\begin{eqnarray}
  \bvec{x}
  =
  \boldsymbol{M}^{-1} \cdot \bvec{Y}
  \ ,
\label{sola}
\end{eqnarray}
and the algorithm is realized by specifying an objective function whose ground 
state is indeed (\ref{sola}). The objective function is not unique, although it must 
be  commensurate with the architecture of the hardware. If the inverse matrix itself 
is required, it can be constructed by solving (\ref{meq}) for each of the $N$ linearly
independent basis vectors for ${\bf Y}$.
It is easy to construct a quadratic objective $H({\bf x})$ whose whose minimum 
is (\ref{sola}), namely
\begin{eqnarray}
  H(\bvec{x})
  &=& 
  \big( \boldsymbol{M}  \bvec{x} - \bvec{Y}  \big)^2 
  =
  \big( \boldsymbol{M}  \bvec{x} - \bvec{Y}  \big)^\smT
  \cdot 
  \big( \boldsymbol{M}  \bvec{x} - \bvec{Y}  \big)
  \label{eq_Hmatrix}
  \ .
\end{eqnarray}
In terms of matrix components, this can be written
\begin{eqnarray}
  H(\bvec{x}) 
  &=& 
  \bvec{x}^\smT \boldsymbol{M}^\smT \boldsymbol{M} \bvec{x} 
  - \bvec{x}^\smT \boldsymbol{M}^\smT \bvec{Y}
  - \bvec{Y}^\smT \boldsymbol{M} \bvec{x} 
  + \bvec{Y}^\smT \bvec{Y} 
  \nonumber \\[5pt]
  &=& 
  \sum_{ijk=1}^N M_{ki} M_{k j} \, x^i x^j 
  - 
  2 \sum_{ij=1}^N Y_j M_{ji} \, x^i 
  + 
  \|\bvec{Y}\|^2
  \ .
\label{mxytwo}
\end{eqnarray}
Note that $\|\bvec{Y}\|^2$ is just a constant, which will not affect the minimization.
In principle all constants can be dropped from the objective function, although we choose
to keep them for completeness. One may obtain a floating point representation of each 
component of \hbox{$\bvec{x} =(x^1, \cdots, x^\smN)^\smT$} by expanding in powers of $2$ 
multiplied by Boolean-valued variables \hbox{$q^i_r \in \{0,1\}$}, 
\begin{eqnarray}
  \chi^i &=& \sum_{r=0}^{R-1} 2^{-r} q^i_r 
\label{chidef}
  \\[5pt]
  x^i &=& 2 \chi^i  - 1
  \ .
\label{xidef}
\end{eqnarray}
As before, the domains are give by $\chi^i \in [0, 2)$ and $x^i \in [-1, 3)$, and upon 
expressing ${\bf x}$ as a function the $q_r^i$ we can recast (\ref{mxytwo}) in the form
%
\begin{eqnarray}
  H[q] 
  &=&
  \sum_{i=1}^N\sum_{r=0}^{R-1} a_r^i \, q^i_r
 + 
  \sum_{i=1}^N \sum_{i \ne j =1}^N \sum_{r=0}^{R-1} \sum_{s = 0}^{R-1} b_{rs}^{ij} \, 
   q^i_r  q^j_s  
  \ .
\label{hsum}
\end{eqnarray}
The coefficients $a_r^i$ are called the {\em weights} and the coefficients $ b_{rs}^{ij}$
are the interaction {\em strengths}. Note that the algorithm requires a connectivity of 
$K_{\rm\scriptscriptstyle NR}$.

Let us first calculate the product $x^i x^j$ in (\ref{mxytwo}). From (\ref{chidef}) and 
(\ref{xidef}) we find 
\begin{eqnarray} 
  x^i x^j 
  &=&  
  \left( 2 \sum_{r=0}^{R-1} 2^{-r} q_r^i - 1 \right) 
  \left( 2 \sum_{r'=0}^{R-1} 2^{-r'} q_{r'}^i - 1 \right) 
  \nonumber 
  \\[5pt]
  &=&
  4\sum_{r r^\prime} 2^{-(r+r^\prime)} q_{r}^i q_{r'}^j - 4\sum_{r} 2^{-r} q_{r}^i  + 1
  \label{xixj_rr}
  \\[5pt]
  &=&
  4\sum_{r \ne r^\prime} 2^{-(r+r^\prime)} q_{r}^i q_{r'}^j 
  +
  4\sum_{r} 2^{-2 r} q_{r}^i
  - 
  4\sum_{r} 2^{-r} q_{r}^i  + 1
  \ ,
\label{xixj}
\end{eqnarray}
where we have used the idempotency condition $(q_r^i)^2 = q_r^i$ in the second 
term of (\ref{xixj}). While the second form is one used by the code, it is more convenient 
algebraically 
to use the first form. Substituting (\ref{xixj_rr}) into the first term in (\ref{mxytwo}) 
gives
\begin{eqnarray} 
  H_1  
  &\equiv& 
  \sum_{ijk} M_{ki} M_{k j} \, x_i x_j 
  \\
  &=&
  \sum_{ijk} M_{ki} M_{k j} \, 
  \left\{ 
  4\sum_{r r^\prime} 2^{-(r+r^\prime)} q_{r}^i q_{r'}^j 
  - 
  4\sum_{r} 2^{-r} q_{r}^i  + 1
   \right\} 
  \\[5pt]
  &=& 
  4 \sum_{ir} \sum_{js} \sum_{k} 2^{-r - s} M_{ki} M_{kj} \, q_{r}^i q_{s}^j 
  - 
  4\sum_{ir}\sum_{k} 2^{-r}  M_{ki} M_{ki} \, q_{r}^i 
  +
    \sum_{ijk} M_{ki} M_{k j} 
  \ .                                
\label{honezero}
\end{eqnarray}
The second term in (\ref{mxytwo}) can be expressed as
\begin{eqnarray}
  H_2 
  &\equiv&
  - 2 \sum_{ij} Y_j M_{ji} \, x_i  
  =
  - 2 \sum_{ij} Y_j M_{ji}  \left( 2 \sum_{r} 2^{-r} q_{r}^i - 1 \right) 
  \\[5pt]
   &=& 
  - 4 \sum_{ij} \sum_{r} 2^{-r} M_{ji} Y_j \, q_{r}^i 
  + 2 \sum_{ij} Y_j M_{ji} 
     \ .
\end{eqnarray}
Adding $H_1$ and $H_2$ gives
\begin{eqnarray}
  H 
  &=&
  4 \sum_{ir} \sum_{js} \sum_{k} 2^{-r - s} M_{ki} M_{kj} \, q_{r}^i q_{s}^j 
  - 
  4\sum_{ir}\sum_{k} 2^{-r}  M_{ki} M_{ki} \, q_{r}^i 
  \\ &&
  - 
  4 \sum_{ij} \sum_{r} 2^{-r} M_{ji} Y_j \, q_{r}^i 
  + 
  2 \sum_{ij} Y_j M_{ji}
  +
 \sum_{ijk} M_{ki} M_{k j}  
  \ .                                
\label{honezero}
\end{eqnarray}
The Ising terms are therefore 
\begin{eqnarray}
  a_r^i &=&  
  4 \cdot 2^{-r} \sum_{k} M_{k i} 
  \left\{  2^{-r} M_{k i}  - \bigg(Y_k + \sum_{j} M_{k j} \bigg) \right\} 
  \label{lair}
  \\[5pt]
  b_{r s}^{ij} 
  &=& 
  4 \cdot 2^{-(r+s)} \sum_{k} M_{k  i}\, M_{k j}
 \ .
\label{lbijrs}
\end{eqnarray}

The physical qubits in are accessed by a $1$-dimensional linear index, while the logical 
qubits in \label{hsum} are defined in terms of the 2-dimensional indicies $i$ and $r$,
where $i = 0, 1, \cdots,  N-1$  and $r = 0, 1, \cdots, R-1$. To map the logical qubits
onto the physical qubits, we re-parameterize the logical qubits by a single index 
$\ell = 0, 1,  \cdots, NR-1$. Now we define a $1$-$1$ mapping between these 
indices and the linear index $\ell = 0, 1, \cdots, N \cdot R - 1$.  This is just an 
ordinary linear indexing for $2$-dimensional matrix elements, so we choose the 
usual row-major linear index mapping,
\begin{eqnarray}
  \ell(i,r) &=& i \cdot R + r  
  \label{elldef}
  \\
  M_{\ell } &=&  M_{ir}
  \ .
  \label{melldef}
\end{eqnarray}
The inverse mapping gives the row and column indices as below,
\begin{eqnarray}
  i_{\ell} &=& \lfloor \ell/R \rfloor  
  \\
  r_{\ell} &=& \ell ~{\rm mod}~ R
  \ ,
\label{rldef}
\end{eqnarray}
where $\lfloor n \rfloor$ is the greatest integer less than or equal to $n$. The 
expression ``$\ell ~{\rm mod}~ R$'' is $\ell$ modulo $R$. This is a $1$-$1$, 
invertible mapping between each pair of values of $i$ and $r$ in the matrix 
index space to every value of $\ell$ in the linear qubits index space. We can 
simply replace sums over all index pairs $i,r$ by a single sum over $\ell$, 
provided we also rewrite any isolated indices in $i$ and $r$ as functions of 
$\ell$ via their inverse mapping.

We may summarize this observation in the following formal identity.
Given some arbitrary quantity, $A$, that depends functionally upon the tuple $(i,r)$, and
possibly upon the individual indices $i$ and $r$,  it is trivial to verify that,
\begin{eqnarray}
  A[(i,r),i,r] = \sum_{\ell=0}^{N \cdot R -1} A[\ell,i_{\ell},r_{\ell}] \, \delta_{i,i_{\ell}} \delta_{rr_{\ell}}
  \ ,
\label{adef}
\end{eqnarray}
where $\ell$, $i_{\ell}$, and $r_{\ell}$ are related as in equations (\ref{elldef})-(\ref{rldef}). This identity is useful
for formal derivations.  For example, we may use it to quickly derive the binary expansion of $x_i$ in 
terms of logical qubits. Inserting (\ref{adef}) into (\ref{xidef}) gives,
\begin{eqnarray}
  x_i &=& 2 \left( \sum_{r=0}^{R-1} 2^{-r} \sum_{\ell=0}^{N \cdot R -1} q_{\ell} \,
            \delta_{i \,i_{\ell}} \delta_{r \, r_{\ell}} \right) - 1 \nonumber 
  \\[5pt]
      &=& 
      2\sum_{\ell=0}^{N \cdot R -1} 2^{-r_{\ell}} \, q_{\ell} \,\delta_{i,i_{\ell}} - 1
  \ .
\label{recon1}
\end{eqnarray}
Clearly, $x_i$ has non-zero contributions only for those indices corresponding to 
$i = i_{\ell} = \lfloor \ell/R \rfloor$,
that is, only from those qubits within a row in the $q^i_r$ array. Also, those contributions 
are summed along that row, {\em i.e.}, over $r_{\ell} = \ell ~{\rm mod}~ R$.
This equation will be used to reconstruct the floating-point solution, $\bvec{x}$, from the
components $q_\ell$ of the binary solution returned from the D-Wave annealing runs. 
The weights and strengths now become
\begin{eqnarray}
  a_{\ell}  
  &=&  
  4 \cdot 2^{-r_{\ell}} \sum_{k} M_{k \, i_{\ell}} 
  \left\{ 2^{-r_{\ell}} M_{k\, i_{\ell}} - (Y_k + \sum_{j} M_{k j} ) \right\} 
  \label{lai}
  \\[5pt]
  b_{\ell m} 
  &=& 
  4 \cdot 2^{-(r_{\ell}+r_{\ell'})} \sum_{k} M_{k \, i_{\ell}}\, M_{k\, i_{m}}
 \ .
\label{lbij}
\end{eqnarray}
For a $2 \times 2$ matrix to 4-bit accuracy, we need $K_{8}$ ($4 \times 2=8$), 
and  to 8-bit accuracy we need $K_{16}$ ($8 \times 2=16$). We have inverted 
matrices up to $3 \times 3$ to 4-bit accuracy, which requires $K_{12}$ ($3 
\times 4 = 12$).  For an $N \times N$ matrix with $R$ bits of resolutions, we 
must construct linear embeddings of $K_{R N}$. We could generalize this 
procedure for complex matrices.

\pagebreak
\section{Calculations}

\subsection{Implementation}
The methods above were implemented using D-Wave's Python SAPI interface and tested
on a large number of floating-point calculations. Initially, we performed floating-point
division on simple test problems with a small resolution. Early on, we discovered 
that larger graph embeddings tended to produce noisier results. To better understand 
what was happening we started with a $K_8$ graph embedding to represent two floating-point 
numbers with only four bits of resolution. Since the D-Wave's dynamic range is limited 
to about a factor of 10 in the scale of the QUBO parameters, we determined that we 
could expect no more than 3 to 4 bits of resolution from any one calculation in any event. 
However, our binary offset representation (\ref{chidef}) implies that we should expect 
no more than 3 bits of resolution in any single run. Indeed, using the $K_8$ embedding, 
we were able to get {\em exact} solutions from the annealer for any division problems 
which had  answers that were multiples of 0.25, between -1.0, and 1.0. Problems in 
this range which had solutions that were {\em not} exact multiples of 0.25 resulted
in approximate solutions, effectively ``rounded'' to the nearest of $\pm 0.25$ or $\pm 0.75$.
At this point we implemented an iterative scheme that uses the current error, or 
residual, as a new input, keeping track of the accumulated floating-point
solution.  

The iteration method has been implemented and tested for floating-point division but
we have not yet implemented iteration for matrix inversion. That can be done
by using the previous residual (error) as the new inhomogeneous term in the
matrix equation. We plan to implement an iterative method in the matrix inversion
code soon. However, we already have good preliminary results on matrix inversion
that encourages us that this should work reasonably well, at least for well-conditioned
matrices. Currently, we are able to solve $2 \times 2$ and $3 \times 3$ linear 
equations involving 
floating-point numbers up to a resolution of 4 bits, and having well-conditioned 
matrices, {\em exactly} for input vectors with elements defined on $[-1,1]$ and 
that are multiples of $0.25$.  Using an example matrix that is poorly-conditioned, 
we find that it is generally not possible to get the right answer without first 
doing some sort of pre-conditioning to the matrix. But more importantly, we were
able to obtain some insight about why ill-conditioned matrices can be difficult
to solve as QUBO problems on a quantum annealer, which gives some hints about how
to ameliorate the problem. We will discuss these results, and the effects of 
ill-conditioning on the QUBO matrix inverse problem below.

\subsubsection{Note on Solution Normalization and Iteration}

Allowing both the division and linear equation QUBO solvers to work for arbitrary 
floating-point numbers, and to allow for iterative techniques, requires normalizing the 
ratio of the current dividend and the divisor, or the residual and matrix, to a value
in a range between $-1$ and $1$. For the division problems, we wanted to avoid 
``dividing in order to divide'', so we normalized each ratio using the difference 
between the binary exponents of $\lfloor \mathrm{divisor} \rfloor$ and $\lfloor 
\mathrm{dividend} \rfloor$. 
These can be found just by using order comparisons, with no explicit divisions. 
Adding $1$ to this yields an ``offset'' - the largest binary exponent of the ratio - 
to within a factor of $2$ ($\pm 1$ in the offset), which is sufficient for scaling our 
QUBO parameters as needed. The fact that our QUBO solutions are always returned in binary 
representation provides a simple way to bound the solution into a range solvable with the 
annealer by simply shifting the binary representation of the current dividend by a few bits
(using the current offset), which is why we refer to the solution exponent as an ``offset''. 
In this way, the solution can easily be guaranteed to be in the correct range without having 
to perform any divisions in Python. The ``offset'' is accumulated and used to construct 
the current approximation to the floating-point solution on each iteration. The iteration 
process continues until the error of the approximate solution is less than 
some tolerance specified by the user.

\subsection{Results for Division}
First, we present some examples for division without iteration. We used a $K_4$ graph 
embedding for expanding the unknown $x$ up to a resolution of 4 bits. However using 
the binary offset representation we can only get a true precisions of 3 bits. 
We solved the simple division problem,
\begin{eqnarray}
  x &=& \frac{y}{m}
  \ .
\label{divprob}
\end{eqnarray}
\subsubsection{Basic Division Solver}
Table~\ref{table_divexact} gives an extensive list of tested exact solutions returned 
by the floating-point annealing algorithm on the D-Wave machine using the $K_4$ graph 
embedding with an effective binary resolution of $3$, corresponding to the multiples
of $0.25$ in the range $[-1,1]$. 
The ``Ground State'' column lists the raw binary vector solutions, corresponding to the 
expansion in Eq (\ref{xdef}).  It is easy to check from Eqs. (\ref{xdef}) and (\ref{chidef_a})
that these give the floating-point solutions found in the corresponding ``D-Wave Solution'' column.  
In all of these cases, values of $\alpha \ge 0.5$ yielded the solution exactly; however,
$\alpha$ is set to 20.0 here because that gives a better approximate solution for the inexact divisions,
and faster convergence for the iterated divisions. It does not change the solutions for the exact
cases. 

\newcommand{\p}{$\phantom{-}$}
\begin{table}[t!]
\caption{\footnoteskip 
  Exact Quantum Annealed Division Problems to $3$-bit Resolution.
}
\begin{tabular}{ |p{2cm}|p{2cm}|p{3cm}||p{3cm}|p{3cm}|p{2cm}|p{2cm}|  }
 \hline
 \multicolumn{7}{|c|}{Division Problems with Exact D-Wave Solutions} \\
 \hline
 y         & m          & x, Exact   & x, D-Wave  & Ground State & Energy    & $\alpha$ \\
 \hline
  $\p1.00$ & $\p1.0$    & $\p1.00$   &  $\p1.00$  & [1,0,0,0]    & -2.0      & ~20.0 \\[-5pt]
  $\p0.50$ & $\p0.5$    & $\p1.00$   &  $\p1.00$  & [1,0,0,0]    & -2.0      & ~20.0 \\[-5pt]
  $\p1.00$ &   -1.0     & -1.00      & -1.00      & [0,0,0,0]    & $\p0.0$   & ~20.0 \\[-5pt]
   -1.00   & $\p1.0$    & -1.00      & -1.00      & [0,0,0,0]    & $\p0.0$   & ~20.0 \\[-5pt]
  $\p0.50$ &   -0.5     & -1.00      & -1.00      & [0,0,0,0]    & $\p0.0$   & ~20.0 \\[-5pt]
    -0.50  & $\p0.5$    & -1.00      & -1.00      & [0,0,0,0]    & $\p0.0$   & ~20.0 \\[-5pt]
  $\p0.75$ & $\p1.0$    & $\p0.75$   &  $\p0.75$  & [0,1,1,1]    & -1.53125  & ~20.0 \\[-5pt]
    -0.75  & $\p1.0$    & -0.75      & -0.75      & [0,0,0,1]    & -0.03125  & ~20.0 \\[-5pt]
  $\p0.75$ &   -1.0     & -0.75      & -0.75      & [0,0,0,1]    & -0.03125  & ~20.0 \\[-5pt]
  $\p0.50$ & $\p1.0$    & $\p0.50$   & $\p0.50$   & [0,1,1,0]    & -1.125    & ~20.0 \\[-5pt]
    -0.50  & $\p1.0$    & -0.50      & -0.50      & [0,0,1,0]    & -0.125    & ~20.0 \\[-5pt]
  $\p0.50$ &   -1.0     & -0.50      & -0.50      & [0,0,1,0]    & -0.125    & ~20.0 \\[-5pt]
  $\p0.25$ & $\p1.0$    & $\p0.25$   & $\p0.25$   & [0,1,0,1]    & -0.78125  & ~20.0 \\[-5pt]
    -0.25  & $\p1.0$    & -0.25      & -0.25      & [0,0,1,1]    & -0.28125  & ~20.0 \\[-5pt]
  $\p0.25$ &   -1.0     & -0.25      & -0.25      & [0,0,1,1]    & -0.28125  & ~20.0 \\[-5pt]
  $\p0.25$ & $\p0.5$    & $\p0.50$   & $\p0.50$   & [0,1,1,0]    & -1.125    & ~20.0 \\[-5pt]
    -0.25  & $\p0.5$    & -0.50      & -0.50      & [0,0,1,0]    & -0.125    & ~20.0 \\[-5pt]
  $\p0.25$ &   -0.5     & -0.50      & -0.50      & [0,0,1,0]    & -0.125    & ~20.0 \\[-5pt]
  $\p0.00$ & $\pm 1.00$ & $\p0.00$   &  $\p0.00$  & [0,1,0,0]    & -0.5      & ~20.0 \\[-5pt]
  $\p0.00$ & $\pm 0.75$ & $\p0.00$   &  $\p0.00$  & [0,1,0,0]    & -0.5      & ~20.0 \\[-5pt]
  $\p0.00$ & $\pm 0.50$ & $\p0.00$   &  $\p0.00$  & [0,1,0,0]    & -0.5      & ~20.0 \\[-5pt]
  $\p0.00$ & $\pm 0.25$ & $\p0.00$   &  $\p0.00$  & [0,1,0,0]    & -0.5      & ~20.0 \\
 \hline                                                                  
\end{tabular}                                                            
\label{table_divexact}                                                   
\end{table}

Table~\ref{table_divapprox} lists some illustrative division problems on $[-1,1]$ 
that do not have solutions which are multiples of $\pm 0.25$, and therefore are
not solved exactly by the quantum annealing algorithm to $3$ bits of resolution.
Note that the energies are different for the ground states because the 
Hamiltonians are somewhat different for these problems. The ``rounding'' 
here occurs naturally in the quantum annealing algorithm as the annealer settles
into the lowest energy ground state that approximates the solution. The last
four problems are ``challenge'' problems for the iterated division solver.

\begin{table}[t!]
\caption{\footnoteskip 
  ``Rounded'' Quantum Annealed Division Solutions to $3$-bit Resolution.
}
\begin{tabular}{ |p{2cm}|p{2cm}|p{3cm}||p{3cm}|p{3cm}|p{2cm}|p{2cm}|  }
 \hline
 \multicolumn{7}{|c|}{Division Problems with Approximate D-Wave Solutions} \\
 \hline
 y         & m    & x, Exact            & x, D-Wave & Ground State & Energy    & $\alpha$ \\
 \hline
  $\p0.90$ & ~1.0 & $\p0.90$            & $\p1.00$  & [1,0,0,0]    & -1.8      & ~20.0 \\[-5pt]
  -0.90    & ~1.0 & -0.90               & -1.00     & [0,0,0,0]    & $\p0.0$   & ~20.0 \\[-5pt]
  $\p0.80$ & ~1.0 & $\p0.80$            & $\p0.75$  & [0,1,0,0]    & -1.6875   & ~20.0 \\[-5pt]
  -0.80    & ~1.0 & -0.80               & -0.75     & [0,0,0,1]    & -0.01875  & ~20.0 \\[-5pt]
  $\p0.70$ & ~1.0 & $\p0.70$            & $\p0.75$  & [0,1,0,0]    & -1.44375  & ~20.0 \\[-5pt]
  -0.70    & ~1.0 & -0.70               & -0.75     & [0,0,0,1]    & -0.04374  & ~20.0 \\[-5pt]
  $\p0.60$ & ~1.0 & $\p0.60$            & $\p0.50$  & [0,1,1,0]    & -1.275    & ~20.0 \\[-5pt]
  -0.60    & ~1.0 & -0.60               & -0.50     & [0,0,1,0]    & -0.075    & ~20.0 \\[-5pt]
  $\p0.40$ & ~1.0 & $\p0.40$            & $\p0.50$  & [0,1,1,0]    & -0.975    & ~20.0 \\[-5pt]
  -0.40    & ~1.0 & -0.40               & -0.50     & [0,0,1,0]    & -0.175    & ~20.0 \\[-5pt]
  $\p0.30$ & ~1.0 & $\p0.30$            & $\p0.25$  & [0,1,0,1]    & -0.84375  & ~20.0 \\[-5pt]
  -0.30    & ~1.0 & -0.30               & -0.25     & [0,0,1,1]    & -0.24375  & ~20.0 \\[-5pt]
  $\p0.20$ & ~1.0 & $\p0.20$            & $\p0.25$  & [0,1,0,1]    & -0.71875  & ~20.0 \\[-5pt]
  -0.20    & ~1.0 & -0.20               & -0.25     & [0,0,1,1]    & -0.31875  & ~20.0 \\[-5pt]
  $\p0.10$ & ~1.0 & $\p0.10$            & $\p0.00$  & [0,1,0,0]    & -0.6      & ~20.0 \\[-5pt]
  -0.10    & ~1.0 & -0.10               & $\p0.00$  & [0,0,1,1]    & -0.4      & ~20.0 \\[-5pt]
  $\p0.30$ & ~0.9 &  $\p0.\bar{3}$      & $\p0.25$  & [0,1,0,1]    & -0.88542  & ~20.0 \\[-5pt]
  -0.30    & ~0.9 & -$0.\bar{3}$        & -0.25     & [0,0,1,1]    & -0.21875  & ~20.0 \\[-5pt]
  $\p1.0$  & ~7.0 &  $\p0.\bar{142875}$ & $\p0.25$  & [0,1,0,1]    & -0.64732  & ~20.0 \\[-5pt]
  -1.0     & ~7.0 & -$0.\bar{142875}$   & -0.25     & [0,0,1,1]    & -0.36161  & ~20.0 \\
 \hline                                                                             
\end{tabular}                                                                       
\label{table_divapprox}
\end{table}
\subsubsection{Iterated Division Solver}
Table~\ref{table_diviter} lists a few example division problems returned from the 
iterated quantum annealing solver. These are problems selected from both 
Tables~\ref{table_divexact} and \ref{table_divapprox} to illustrate the nature
of the solutions returned for both cases. These problems were iterated to an
error tolerance of $1.0 \times 10^{-6}$. The four ``challenge'' problems from 
Table~\ref{table_divapprox} can now be solved with the iterative method. The ground 
state is no longer given since the solution is generally the concatenation of multiple 
binary vectors for every iteration. Instead, the number of iterations is listed 
in the last column. Note that some of the energies are the same for the solutions 
of different problems.  We have also left out an ``Energy'' column, since it only 
was calculated for the partial solution from the last iteration. 

\begin{table}[t!]
\caption{\footnoteskip 
  Iterated Quantum Annealed Division Problems to Resolution $1.0 \times 10^{-6}$.
}
\begin{tabular}{ |p{2cm}|p{2cm}|p{3cm}||p{3cm}|p{3cm}|p{2cm}|p{2cm}|  }
 \hline
 \multicolumn{6}{|c|}{Iterated Division Problems on the D-Wave Annealer} \\
 \hline
 y         & m     & x, Exact            & x, D-Wave      & $\alpha$  & Iterations \\
 \hline
  $\p0.25$ &  ~1.0 & $\p0.25$            & $\p0.25$       & ~20.0  & ~1 \\ [-5pt]   
  -0.25    &  ~1.0 & -0.25               & -0.25          & ~20.0  & ~1 \\ [-5pt]   
  $\p0.50$ &  ~1.0 & $\p0.50$            & $\p0.50$       & ~20.0  & ~1 \\ [-5pt]   
  -0.50    &  ~1.0 & -0.50               & -0.50          & ~20.0  & ~1 \\ [-5pt]   
  $\p0.75$ &  ~1.0 & $\p0.75$            & $\p0.75$       & ~20.0  & ~1 \\ [-5pt]   
  -0.75    &  ~1.0 & -0.75               & -0.75          & ~20.0  & ~1 \\ [-5pt]   
  $\p0.80$ &  ~1.0 & $\p0.80$            &  $\p0.799999$  & ~20.0  & ~5 \\ [-5pt]   
  -0.80    &  ~1.0 & -0.80               & -0.799999      & ~20.0  & ~5 \\ [-5pt]   
  $\p0.70$ &  ~1.0 & $\p0.70$            & $\p0.700000$   & ~20.0  & ~5 \\ [-5pt]   
  -0.70    &  ~1.0 & -0.70               & -0.700000      & ~20.0  & ~5 \\ [-5pt]   
  $\p0.10$ &  ~1.0 & $\p0.10$            & $\p0.999999$   & ~20.0  & ~5 \\ [-5pt]   
  -0.10    &  ~1.0 & -0.10               & -0.999999      & ~20.0  & ~5 \\ [-5pt]   
  $\p0.30$ &  ~0.9 & $\p0.\bar{3}$       & $\p0.333333$   & ~20.0  & ~10 \\[-5pt]   
  -0.30    &  ~0.9 & -$0.\bar{3}$        & -0.333333      & ~20.0  & ~10 \\[-5pt]   
  $\p1.0$  &  ~7.0 &  $\p0.\bar{142875}$ &  $\p0.1248751$ & ~20.0  & ~7 \\ [-5pt]   
  -1.0     &  ~7.0 & -$0.\bar{142875}$   & -0.1248751     & ~20.0  & ~7 \\ 
 \hline                                                          
\end{tabular}                                                    
\label{table_diviter}                                            
\end{table}                                                      

\subsection{Results for Matrix Equations}
Note that we have occasionally been somewhat loose in calling this ``matrix inversion'', since 
we are technically solving the linear equations, rather than directly inverting the matrices.
However, for the problems considered here, we may simply obtain the solutions to the 
equations using trivial orthonormal eigenvectors such as $(1,0)$ and $(0,1)$, in which case 
the inverse of the matrix will just be the matrix having those solutions as columns.

The linear equation algorithm was implemented and used to solve several $2 \times 2$ and
$3 \times 3$ matrices on the D-Wave quantum annealer. Floating-point numbers are represented
using the same offset binary representation as was used for the division problems. Thus,
there are $4$ qubits per floating-point number. As in the previous section, this gives an 
effective resolution of $3$ bits for floating-point numbers defined on $[-1,1]$. In this case,
however, we employed the normalization technique discussed in the the division iteration to 
allow solutions with positive and negative floating-point numbers with larger magnitudes than $1$. 
But, in these problems we still use solution values with relatively small magnitudes, and 
within an order of magnitude of each other for all solution vector elements. All of 
the cases shown here are matrix equations with exact solutions, in which case the values 
of the solution vector elements are multiples of $0.25$. This suggests that we could have 
the iterative solver for the matrix inversion working very soon.

In general, every qubit constituting a floating-point number may be coupled
to every other qubit for the same number. In turn, every logical qubit may be connected 
to every other logical qubit, which implies that every qubit in the logical qubit 
representation of the problem, may be coupled to every other logical qubit in the 
problem. Therefore, the linear solution algorithm is implemented using a $K_{8}$ graph 
embedding to solve $2 \times 2$ matrix equations, having a $2$ dimensional solution vector 
with $4$ qubits per element, and using a $K_{12}$ graph embedding to solve $3 \times 3$
matrix equations, having a $3$ dimensional solution vector with $4$ qubits per element.

Most of these solutions involve well-conditioned matrices; however, one does not generally
find a feasible solution when using an ill-conditioned matrix. This is illustrated 
in two cases, one with a of a $2 \times 2$ matrix another with a $3 \times 3$ matrix.  
We were able to obtain the correct solutions by pre-conditioning these matrices before 
converting to QUBO form, however the $3 \times 3$ matrix, still had a nearly degenerate 
ground state and required a very large chaining penalty $\alpha$ to get the correct solution.  
This is analyzed and discussed in detail below.

\subsubsection{Simple Analytic Problem}
Recalling equation (\ref{meq}), we shall obtain solutions $\bvec{x}$ of the 
following matrix equation,
\begin{eqnarray}
  \boldsymbol{M} \cdot \bvec{x} =  \bvec{Y},
  \ 
\end{eqnarray}
using values of $\boldsymbol{M}$ and $\bvec{Y}$ listed in the Appendix. Here 
we present the first two tests as an example. Consider the following matrix,

\begin{eqnarray}
  \boldsymbol{M} = \left(
    \begin{array}{cc}
      \frac{1}{2} & \frac{3}{2} \\
      \frac{3}{2} & \frac{1}{2} 
    \end{array} \right).
\end{eqnarray}
We can solve equation (\ref{meq}) for $\boldsymbol{M}$, with the following
two $Y$ vectors,
\begin{eqnarray}
  \bvec{Y_1} = \left(
    \begin{array}{c}
      1 \\ 
      0  
    \end{array} \right),
  & 
  \bvec{Y_2} = \left(
    \begin{array}{c}
      0 \\ 
      1  
    \end{array} \right).
\end{eqnarray}
The exact solutions are,
\begin{eqnarray}
  \bvec{x_1} = \left(
    \begin{array}{c}
      -\frac{1}{4} \\ 
      ~\frac{3}{4}  
    \end{array} \right),
  &
  \bvec{x_2} = \left(
    \begin{array}{c}
     ~\frac{3}{4} \\ 
     -\frac{1}{4}
    \end{array} \right).
\end{eqnarray}
We may obtain $M^{-1}$ simply as,
\begin{eqnarray}
  \boldsymbol{M^{-1}} = \left(
    \begin{array}{cc}
      -\frac{1}{4} &  ~\frac{3}{4} \\
      ~\frac{3}{4} & -\frac{1}{4} 
    \end{array} \right)  
\end{eqnarray}
In the next section we summarize all of the solutions obtained by the DWave for all
of our test problems.

\subsubsection{QUBO Solution Results}
The solutions for the $2 \times 2$ linear solves are presented in Table~\ref{table_matsolves_one}.
Notice that all of the test problems are presented with $\alpha=20.0$ except for the last two.  This was 
done to illustrate the affect of pre-conditioning for the ill-conditioned case.  However, for this example,
the difference disappeared above $\alpha=2.0$, and both began to give incorrect answers below $\alpha=1.5$. 
This is in contrast to the $3 \times 3$ matrix solution cases, which are evidently more sensitive to 
condition number than the $2 \times 2$ tests.

\begin{table}[hb]
\caption{\footnoteskip 
  $2 \times 2$  Matrix Equation Solutions to $3$-bit Resolution.
}
\begin{tabular}{ |p{1cm}|p{2.5cm}||p{3cm}|p{3cm}|p{2cm}|p{1cm}|  }
 \hline
 \multicolumn{6}{|c|}{Division Problems with Approximate D-Wave Solutions} \\
 \hline
 Test      & Exact Solution      & D-Wave Solution    & Ground State        & Energy       & $\alpha$ \\
 \hline
  $1(a)$   & $( -0.25,\p0.75)$   & $( -0.25,\p0.75)$  & [0,0,1,1,0,1,1,1]   & $-2.167$     & ~20.0 \\[-5pt]
  $1(b)$   & $(\p0.75, -0.25)$   & $(\p0.75, -0.25)$  & [0,1,1,1,0,0,1,1]   & $-2.167$     & ~20.0 \\[-5pt]
  $1(c)$   & $(\p1.00,\p1.00)$   & $(\p1.00,\p1.00)$  & [1,0,0,0,1,0,0,0]   & $-0.444$     & ~20.0 \\[-5pt]
  $1(d)$   & $( -1.00,\p1.00)$   & $( -1.00,\p1.00)$  & [0,0,0,0,1,0,0,0]   & $-1.889$     & ~20.0 \\[-5pt]
  $1(e)$   & $(\p1.00, -1.00)$   & $(\p1.00, -1.00)$  & [1,0,0,0,0,0,0,0]   & $-1.650$     & ~20.0 \\[-5pt] 
  $1(f)$   & $(\p1.00,\p0.00)$   & $(\p1.00,\p0.00)$  & [0,0,0,0,0,1,0,0]   & $-2.125$     & ~20.0 \\[-5pt]  
  $1(g)$   & $(\p0.25, -0.50)$   & $(\p0.25, -0.50)$  & [0,1,0,1,0,0,1,0]   & $-0.925$     & ~20.0 \\[-5pt]  
  $1(h)$   & $(\p0.25,\p0.25)$   & $(\p0.25,\p0.25)$  & [0,1,0,1,0,1,0,1]   & $-2.03125$   & ~20.0 \\[-5pt]  
  $1(i)$   & $(\p2.00,\p1.00)$   & $(\p2.00,\p1.00)$  & [1,1,0,0,1,0,0,0]   & $-2.450126$  & ~20.0 \\[-5pt]   
  $1(j)$   & $(\p2.00,\p1.00)$   & $(\p2.00,\p1.00)$  & [1,1,0,0,1,0,0,0]   & $-2.532545$  & ~20.0 \\[-5pt]
  $1(i)$   & $(\p2.00,\p1.00)$   & $(\p2.50,\p0.75)$  & [1,1,1,0,0,1,1,1]   & $-2.887689$  & ~1.5 \\[-5pt]   
  $1(j)$   & $(\p2.00,\p1.00)$   & $(\p2.00,\p1.00)$  & [1,1,0,0,1,0,0,0]   & $-2.951557$  & ~1.75 \\
 \hline                                                                             
\end{tabular}                                                                       
\label{table_matsolves_one}
\end{table}

The $3 \times 3$ matrix solutions are presented in Table~\ref{table_matsolves_two}. Note that We have 
not included the 12 digit binary ground states here because they take up too much room in the table and 
are not particularly illuminating.
Problems $2(f)$ and $2(g)$ are the ill-conditioned matrix test and its pre-conditioned equivalent. 
For $\alpha=20.0$ both versions of the poorly-conditioned problem gave only D-Wave solutions with 
broken chains. One only begins to get solutions with unbroken chains at a value of $\alpha$ above $1000$, 
but those solutions are generally wrong and basically random until one gets to a very high $\alpha$.  
We discuss this in greater detail in the following section.
\begin{table}[hb]
\caption{\footnoteskip 
  $3 \times 3$  Matrix Equation Solutions to $3$-bit Resolution.
}
\begin{tabular}{ |p{1cm}|p{3.5cm}||p{3.5cm}|p{3cm}|p{2cm}|  }
 \hline
 \multicolumn{5}{|c|}{Division Problems with Approximate D-Wave Solutions} \\
 \hline
 Test      & Exact Solution      & D-Wave Solution     & Energy       & $\alpha$ \\
 \hline
  $2(a)$   & $(\p0.25,-0.5,\p1.0)$   & $(\p0.25,-0.5,\p1.0)$  & $-15.5625$  & ~20.0 \\[-5pt]
  $2(b)$   & $(\p0.25,-0.5,\p0.0)$   & $(\p0.25,-0.5,\p0.0)$  & $-12.5625$  & ~20.0 \\[-5pt]
  $2(c)$   & $(\p0.25,\p0.0,-0.5)$   & $(\p0.25,\p0.0,-0.5)$  & $-13.5$     & ~20.0 \\[-5pt]
  $2(d)$   & $(\p1.0,\p0.25,-0.5)$   & $(\p1.0,\p0.25,-0.5)$  & $-15.6875$  & ~20.0 \\[-5pt]
  $2(e)$   & $(\p0.0,\p0.25,-0.5)$   & $(\p0.0,\p0.25,-0.5)$  & $-12.75$    & ~20.0 \\[-5pt] 
  $2(f)$   & $(\p0.0,\p0.25,-0.75)$  & ~broken chains         & ~N/A        & ~20.0 \\[-5pt]  
  $2(g)$   & $(\p0.0,\p0.25,-0.75)$  & ~broken chains         & ~N/A        & ~20.0 \\[-5pt]  
  $2(f)$   & $(\p0.0,\p0.25,-0.75)$  & $(\p1.75,\p1.25,0.75)$ & $-58.188$   & ~2200.0 \\[-5pt]  
  $2(g)$   & $(\p0.0,\p0.25,-0.75)$  & $(\p0.0,\p0.25,-0.75)$ & $-557.437$  & ~2200.0 \\ 
 \hline                                                                             
\end{tabular}                                                                       
\label{table_matsolves_two}
\end{table}

\subsection{Discussion}
The algorithms described here generally worked quite well for these small test cases with the 
exception of the ill-conditioned $3 \times 3$ matrix. The ill-conditioned cases clearly demonstrate not 
only the limitations of quantum annealing applied solving linear equations, but the limitations of quantum
annealing, in general.  
Consider the two ill-conditioned tests presented here. When translated  to a QUBO problem, the Hamiltonian
spectra for these tests contain many energy eigenvalues very close to the ground state energy. When these
are embedded within a larger graph of physical qubits they result in a very nearly degenerate ground 
state, typically with thousands  of states having energies within the energy uncertainty of the 
ground state over the annealing time, $\tau$, given by

\begin{eqnarray}
  \Delta E = \frac{\hbar}{\tau}
  \ .
\end{eqnarray}

Consider a set of excited states with energy, $E_n$ for $n > 0$, with $n=0$, corresponding to the ground state with
energy denoted by $E_0$, and with $E_n$ ordered by energy. The quantum annealer near the ground state evolves 
adiabatically whenever 
\begin{eqnarray}
  E_1-E_0 \gg \frac{\hbar}{\tau}
  \ 
\end{eqnarray}
This is the adiabatic condition for quantum time evolution\cite{messiah}.  However, when this condition 
is badly violated, which can occur dynamically since the instantaneous energies (eigenvalues of H) are
time dependent, the time evolution for the system near the ground state deviates significantly from 
adiabatic behavior, resulting in a relatively slowly evolving superposition of all the eigenstates states 
that are close in energy to the ground state. Now, the energy spectra corresponding to poorly conditioned matrices 
have a large number of eigenstates sufficiently near the ground state to strongly violate the adiabaticity 
condition. Furthermore, these states, in general, will have no correlation to the solution encodings
for any particular problem (e.g., the offset binary floating-point representation). For example, 
they are not, in general, related in any meaningful way to Hamming distance.  Therefore, these problems
effectively cannot resolve the true ground state and tend to give nearly random lowest energy "solutions" 
when the final state is measured on any individual annealing run. Since there are so many of these states 
for ill-conditioned problems, a very large number of "reads" (individual runs) may be have to be specified 
to sufficient sample the solution space to find the true ground state.

The behavior of the QUBO energies for the ill-conditioned matrices is illustrated in the figures below. 
The final states are binary vectors corresponding to binary numbers up to $2^N$ for $N$ qubits.
Since our test problems were reasonably small, we directly computed the energies for all $256$ 
final states for the $2 \times 2$ matrices, and all $4096$ states for the $3 \times 3$ matrices.  
The array of binary solution states were plotted as what we call ``Gray'' projections. By this
we mean that the energy is plotted against the solution number ordered in a Gray code sequence.  
Gray codes are cyclically ordered lists of binary numbers such that every consecutive pair of numbers differ 
in only one bit, i.e., there us a Hamming distance of $1$ between any two sequential numbers\cite{gray_code}. 
We employed the standard ``reflected binary'' Gray code which is conveniently generated by a recursive 
algorithm. Table~\ref{table_gray_code} shows an example Gray code for the case of 4 bits.
\begin{table}[hb]
\caption{\footnoteskip
  Example Gray code for 4 bits
}
\begin{tabular}{ |p{1cm}|p{3cm}|p{3cm}|  }
 \hline
 \multicolumn{3}{|c|}{4 Bit Reflected Binary Gray Code} \\
 \hline
 N      & Standard Binary     & Gray Code  \\
 \hline
  $0$   & $0000$              & $0000$ \\[-5pt]
  $1$   & $0001$              & $0001$ \\[-5pt]
  $2$   & $0010$              & $0011$ \\[-5pt]
  $3$   & $0011$              & $0010$ \\[-5pt]
  $4$   & $0100$              & $0110$ \\[-5pt] 
  $5$   & $0101$              & $0111$ \\[-5pt] 
  $6$   & $0110$              & $0101$ \\[-5pt] 
  $7$   & $0111$              & $0100$ \\[-5pt]   
  $8$   & $1000$              & $1100$ \\[-5pt]
  $9$   & $1001$              & $1101$ \\[-5pt]
  $10$  & $1010$              & $1111$ \\[-5pt]
  $11$  & $1011$              & $1110$ \\[-5pt]
  $12$  & $1100$              & $1010$ \\[-5pt] 
  $13$  & $1101$              & $1011$ \\[-5pt] 
  $14$  & $1110$              & $1001$ \\[-5pt] 
  $15$  & $1111$              & $1000$ \\
 \hline                                                                             
\end{tabular}                                                                       
\label{table_gray_code}
\end{table}
Gray code ordering was used for these plots because, although it is impossible to represent all nearest-neighbor 
connections for $K_8$ and $K_{12}$ graphs in a $1$ dimensional line plot, we were able to at least ensure 
that all of the adjacent values in the energy plots correspond to neighboring states, which gives some 
information about the local energy landscape, such as whether neighboring states are close in energy, 
indicating a relatively ``flat'' displacement {\em along that direction}, or if there is a large jump, indicating a 
steep ``cliff'' or ``valley'' in the energy landscape. Note, also, that this Gray code sequence has global permutation 
symmetry among its digits, and a cyclic symmetry that repeats periodically for every $2^2$ numbers in each digit place. 
Since this periodicity is also reflected in some periodicity of states separated by a Hamming distance of $1$, this 
makes it possible to infer features of the local energy surface along other directions such as the existence of very 
deep, but nearly flat, ``canyon floors'' in the global energy surface. One can see this by noting periodic behavior 
with even periodicity in the $1$ dimensional energy plots. 

\begin{figure}[h!]
\includegraphics[scale=0.45]{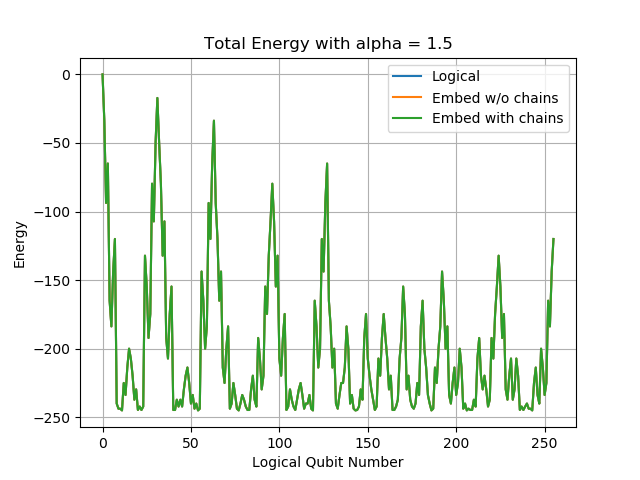} 
\caption{\footnoteskip  
Gray Projection of Energy vs State for Test 1(i)
}
\label{fig_twobytwo_ill_cond}
\end{figure}

\begin{figure}[h!]
\includegraphics[scale=0.45]{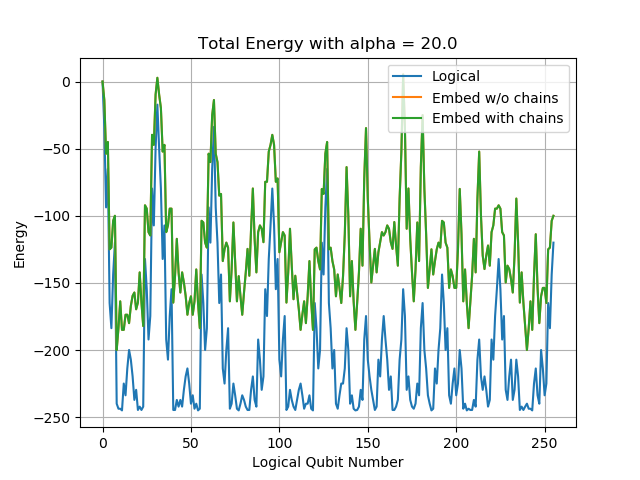} 
\caption{\footnoteskip  
Gray Projection of Energy vs State for Test 1(i) without Weighted Counter-terms
}
\label{fig_twobytwo_ill_condtwo}
\end{figure}

\begin{figure}[h!]
\includegraphics[scale=0.45]{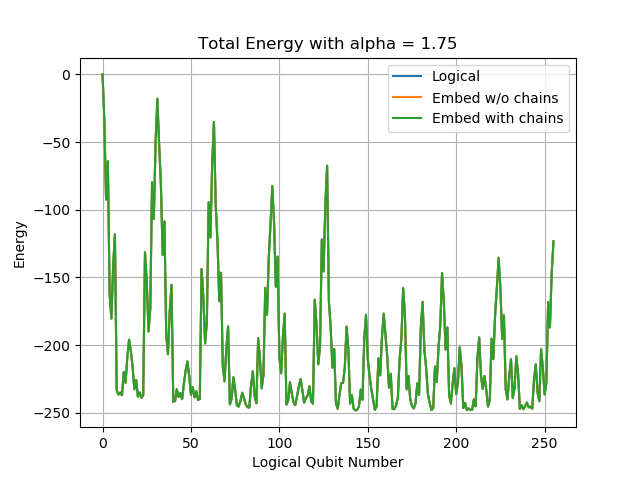} 
\caption{\footnoteskip  
Gray Projection of Energy vs State for Test 1(j)
}
\label{fig_twobytwo_pre_cond}
\end{figure}

Figure~\ref{fig_twobytwo_ill_cond} shows the energy surface for the $2 \times 2$ ill-conditioned matrix. 
Note the denseness of the energy spectrum near the ground state energy.  The alpha value is listed on 
this plot, however, with our counter-term parameter setting strategy the energy of all the 
final states (when there are no broken chains) is actually independent of $\alpha$.
To illustrate this point, compare this to Figure~\ref{fig_twobytwo_ill_condtwo}, which is 
the same test problem but computed {\em without} using the counter-terms weighted to subtract out the chaining 
penalty energy contribution. This plot clearly shows why it is crucial to remove the error in the energy 
caused by not accounting for varying different chain sizes when applying the chaining penalty. Note from the 
plot that this error is inhomogeneous in state space. Therefore, the lowest energy state for the embedded Hamiltonian 
will frequently not be the correct one corresponding to the logical Hamiltonian.  Note, also, that when the energy of all 
the chained logical qubits is restricted to be $0$, the plots for the embedded and logical QUBO energies overlay each other
exactly. Figure~\ref{fig_twobytwo_pre_cond} shows the energy for the  pre-conditioned version of this problem 
with the counter-term applied.  Note that the pre-conditioning has moved many of the previous low-energy eigenvalues
significantly higher.

\begin{figure}[t!]
\includegraphics[scale=0.45]{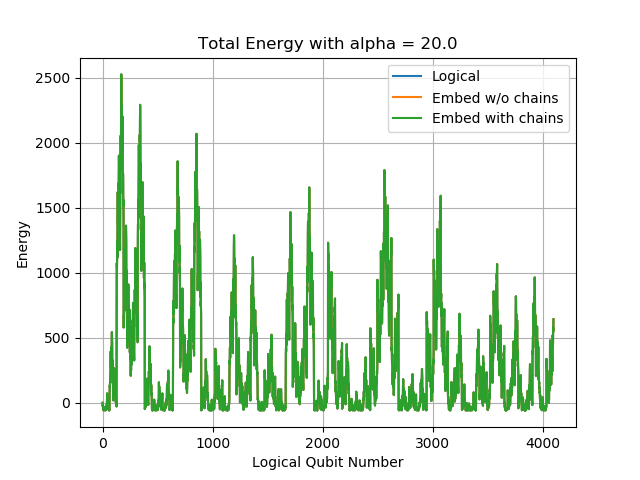} 
\caption{\footnoteskip  
Gray Projection of Energy vs State for Test 2(f)
}
\label{fig_threebythree_ill_cond}
\end{figure}
\begin{figure}[t!]
\includegraphics[scale=0.45]{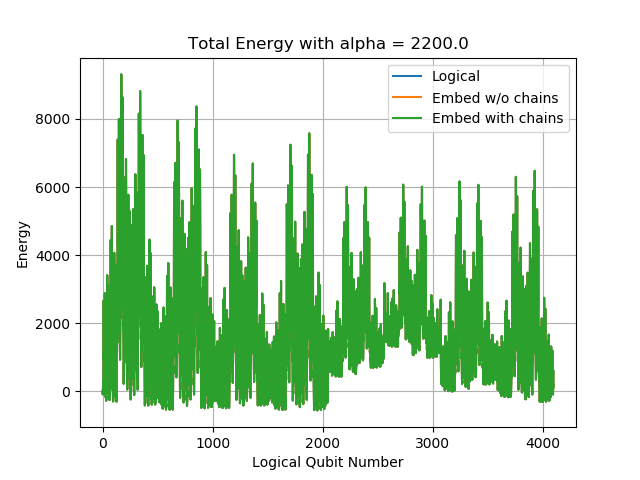} 
\caption{\footnoteskip  
Gray Projection of Energy vs State for Test 2(g)
}
\label{fig_threebythree_pre_cond}
\end{figure}

The $3 \times 3$ ill-conditioned matrix was much more problematic. This is partly because of the nature
of this test matrix, which is very nearly singular with an approximate $2$ dimensional kernel space. The energy spectrum
of the ill-condition problem is given in Figure~\ref{fig_threebythree_ill_cond}, and the energy spectrum 
of the pre-conditoned version is given in Figure~\ref{fig_threebythree_pre_cond}.   Although the 
pre-conditioned version is significantly better, it still required a very large chaining penalty, and 
even with that, it required thousands of ``reads'' (annealing runs) to reliably obtain the correct answer. 
In this case, it required sampling the state space with as many as $2500$ or more ``reads'' with $20 \mu s$ anneal times. 
Only several orders of magnitude longer anneal times began to allow for a smaller number of reads. Here there 
were $2^{12}=4096$ final states for the logical Hamiltonian, so we were actually sampling with more than 
half as many distinct runs as the number of possible states. However, the $K_{12}$ embedding used for 
these runs had $56$ qubits. Therefore, the total state space for all qubits in the embedding was of 
size $2^{56} \approx 7 \times 10^{16}$, so perhaps this large sampling requirement is not as bad as it
may at first appear. Still, it is clear that there are still far too many excited states too close to the
ground state and that the system is far from perfectly adiabatic here. However, it should be noted that 
the pre-conditioning still improved things somewhat, because we could not get the correct answer with 
even $10000$ ``reads'' for the original ill-conditioned problems.

The pre-conditioning method we used was very crude, rather ad-hoc, and ill-suited to practical
matrix inversion problems for classical algorithms, but the intent was simply to test the effects 
of a simple pre-conditioning on the quantum annealing solutions. We have been studying this
issue and believe it may be possible to pre-condition the problems better for solving these
problems on a quantum annealing machine. In fact, we suspect that methods related to this approach
may allow other QUBO problems suffering from similar pathologies to be ``pre-conditioned''
in way that better separates the ground state energy and allows more practical solutions
on the annealer.  This, however, is still work in progress, and we plan to further develop
and test those ideas in the near future.

\pagebreak
\clearpage
\appendix
\section{Matrix Test Problems}
The problems we solved to test our quantum annealing algorithm to solve equation
(\ref{meq}) are listed below. Note that, although the QUBO $B_{ij}$ matrix is symmetric 
by construction, the matrix $M$ need not be symmetric.

\begin{enumerate}
  \item Test Problems with $2 \times 2$ Matrices
    \begin{enumerate}
    \item[] Test $1(a)$: 
      \begin{eqnarray}
        \boldsymbol{M} = \left( 
          \begin{array}{cc}
            0.5 & 1.5 \\
            1.5 & 0.5 
          \end{array} \right),
        & 
        \bvec{Y} = \left(
          \begin{array}{c}
            1.0 \\ 
            0.0  
          \end{array} \right), 
        & 
        \bvec{x} = \left(
          \begin{array}{c}
            -0.25 \\ 
            0.75  
          \end{array} \right) 
        \notag
      \end{eqnarray}
    \item[] Test $1(b)$: 
      \begin{eqnarray}
        \boldsymbol{M} = \left( 
          \begin{array}{cc}
            0.5 & 1.5 \\
            1.5 & 0.5 
          \end{array} \right),
        & 
        \bvec{Y} = \left(
          \begin{array}{c}
            0.0 \\ 
            1.0  
          \end{array} \right),       
        & 
        \bvec{x} = \left(
          \begin{array}{c}
            0.75 \\ 
            -0.25  
          \end{array} \right) 
        \notag
      \end{eqnarray}
    \item[] Test $1(c)$: 
      \begin{eqnarray}
        \boldsymbol{M} = \left( 
          \begin{array}{cc}
            2.0 & -1.0 \\
            -0.5 &  0.5 
          \end{array} \right),
        & 
        \bvec{Y} = \left(
          \begin{array}{c}
            1.0 \\ 
            0.0  
          \end{array} \right),       
        &
        \bvec{x} = \left(
          \begin{array}{c}
            1.0 \\ 
            1.0  
          \end{array} \right) 
        \notag
      \end{eqnarray}
    \item[] Test $1(d)$: 
      \begin{eqnarray}
        \boldsymbol{M} = \left( 
          \begin{array}{cc}
            1.0 & 2.0 \\
            0.5 & 0.5 
          \end{array} \right),
        & 
        \bvec{Y} = \left(
          \begin{array}{c}
            1.0 \\ 
            0.0  
          \end{array} \right),       
        &
        \bvec{x} = \left(
          \begin{array}{c}
            -1.0 \\ 
            1.0  
          \end{array} \right) 
        \notag
      \end{eqnarray}
    \item[] Test $1(e)$: 
      \begin{eqnarray}
        \boldsymbol{M} = \left( 
          \begin{array}{cc}
            3.0 & 2.0 \\
            2.0 & 1.0 
          \end{array} \right),
        & 
        \bvec{Y} = \left(
          \begin{array}{c}
            1.0 \\ 
            1.0  
          \end{array} \right),       
        &
        \bvec{x} = \left(
          \begin{array}{c}
            1.0 \\ 
            -1.0  
          \end{array} \right) 
        \notag
      \end{eqnarray}
    \item[] Test $1(f)$: 
      \begin{eqnarray}
        \boldsymbol{M} = \left( 
          \begin{array}{cc}
            1.0 &  0.5 \\
            1.0 & -0.5 
          \end{array} \right),
        & 
        \bvec{Y} = \left(
          \begin{array}{c}
            1.0 \\ 
            1.0  
          \end{array} \right),      
        &
        \bvec{x} = \left(
          \begin{array}{c}
            1.0 \\ 
            0.0  
          \end{array} \right) 
        \notag
      \end{eqnarray}
    \item[] Test $1(g)$: 
      \begin{eqnarray}
        \boldsymbol{M} = \left( 
          \begin{array}{cc}
            0.0 & -2.0 \\
           -2.0 & -1.5 
          \end{array} \right),
        & 
        \bvec{Y} = \left(
          \begin{array}{c}
            1.0 \\ 
            0.25  
          \end{array} \right),      
        &
        \bvec{x} = \left(
          \begin{array}{c}
            0.25 \\ 
           -0.5  
          \end{array} \right) 
        \notag
      \end{eqnarray}
    \item[] Test $1(h)$: 
      \begin{eqnarray}
        \boldsymbol{M} = \left( 
          \begin{array}{cc}
            0.0 & -2.0 \\
           -2.0 & -1.5 
          \end{array} \right),
        & 
        \bvec{Y} = \left(
          \begin{array}{c}
           -0.5 \\ 
           -0.875  
          \end{array} \right),      
        &
        \bvec{x} = \left(
          \begin{array}{c}
            0.25 \\ 
            0.25  
          \end{array} \right) 
        \notag
      \end{eqnarray}
    \item[] Test $1(i)$: Ill-conditioned problem with $\kappa \approx 25$ 
      \begin{eqnarray}
        \boldsymbol{M} = \left( 
          \begin{array}{cc}
            1.0 & 2.0   \\
            2.0 & 3.999 
          \end{array} \right),
        & 
        \bvec{Y} = \left(
          \begin{array}{c}
            4.0    \\ 
            7.999  
          \end{array} \right),      
        &
        \bvec{x} = \left(
          \begin{array}{c}
            2.0 \\ 
            1.0  
          \end{array} \right) 
        \notag
      \end{eqnarray}
    \item[] Test $1(j)$: Pre-conditioned version of $1(i)$ with $\kappa = 5.0$ 
      \begin{eqnarray}
        \boldsymbol{M} = \left( 
          \begin{array}{cc}
            1.80026 & 1.6019 \\
            1.6019 & 4.19974 
          \end{array} \right),
        & 
        \bvec{Y} = \left(
          \begin{array}{c}
            5.2007  \\ 
	    7.40013  
          \end{array} \right),      
        &
        \bvec{x} = \left(
          \begin{array}{c}
            2.0 \\ 
            1.0  
          \end{array} \right) 
        \notag
      \end{eqnarray}
    \end{enumerate}

  \item Matrix Problems with $3 \times 3$ Matrices
    \begin{enumerate}
    \item[] Test $2(a)$: 
      \begin{eqnarray}
        \boldsymbol{M} = \left( 
          \begin{array}{ccc}
            0.0 & -2.0 & 0.0 \\
           -2.0 &  1.5 & 0.0 \\
            0.0 &  0.0 & 1.0
          \end{array} \right),
        & 
        \bvec{Y} = \left(
          \begin{array}{c}
            1.0   \\ 
            0.25  \\
            1.0
          \end{array} \right), 
        & 
        \bvec{x} = \left(
          \begin{array}{c}
            0.25 \\ 
           -0.5  \\
            1.0
          \end{array} \right) 
        \notag
      \end{eqnarray}
    \item[] Test $2(b)$: 
      \begin{eqnarray}
        \boldsymbol{M} = \left( 
          \begin{array}{ccc}
            0.0 & -2.0 & 0.0 \\
           -2.0 &  1.5 & 0.0 \\
            0.0 &  0.0 & 1.0
          \end{array} \right),
        & 
        \bvec{Y} = \left(
          \begin{array}{c}
            1.0   \\ 
            0.25  \\
            0.0
          \end{array} \right), 
        & 
        \bvec{x} = \left(
          \begin{array}{c}
            0.25 \\ 
           -0.5  \\
            0.0
          \end{array} \right) 
        \notag
      \end{eqnarray}
    \item[] Test $2(c)$: 
      \begin{eqnarray}
        \boldsymbol{M} = \left( 
          \begin{array}{ccc}
            1.0 &  0.0 &  0.0  \\
            0.0 &  0.0 & -2.0 \\
            0.0 & -2.0 & -1.5
          \end{array} \right),
        & 
        \bvec{Y} = \left(
          \begin{array}{c}
            1.0   \\ 
            0.0   \\
            0.25
          \end{array} \right), 
        & 
        \bvec{x} = \left(
          \begin{array}{c}
            0.25 \\ 
            0.0  \\
           -0.5
          \end{array} \right) 
        \notag
      \end{eqnarray}
    \item[] Test $2(d)$: 
      \begin{eqnarray}
        \boldsymbol{M} = \left( 
          \begin{array}{ccc}
            1.0 &  0.0 &  0.0  \\
            0.0 &  0.0 & -2.0 \\
            0.0 & -2.0 & -1.5
          \end{array} \right),
        & 
        \bvec{Y} = \left(
          \begin{array}{c}
            1.0   \\ 
            1.0  \\
            0.25
          \end{array} \right), 
        & 
        \bvec{x} = \left(
          \begin{array}{c}
            1.0 \\ 
            0.25  \\
           -0.5
          \end{array} \right) 
        \notag
      \end{eqnarray}
    \item[] Test $2(e)$: 
      \begin{eqnarray}
        \boldsymbol{M} = \left( 
          \begin{array}{ccc}
            1.0 &  0.0 &  0.0  \\
            0.0 &  0.0 & -2.0 \\
            0.0 & -2.0 & -1.5
          \end{array} \right),
        & 
        \bvec{Y} = \left(
          \begin{array}{c}
            0.0   \\ 
            1.0  \\
            0.25
          \end{array} \right), 
        & 
        \bvec{x} = \left(
          \begin{array}{c}
            0.0 \\ 
            0.25  \\
           -0.5
          \end{array} \right) 
        \notag
      \end{eqnarray}
    \item[] Test $2(f)$: Ill-conditioned problem with $\kappa \approx 78$ 
      \begin{eqnarray}
        \boldsymbol{M} = \left( 
          \begin{array}{ccc}
            -4.0 &   6.0  &  1.0 \\ 
             8.0 & -11.0  & -2.0 \\  
            -3.0 &   4.0  &  1.0     
          \end{array} \right),
        & 
        \bvec{Y} = \left(
          \begin{array}{c}
            0.75 \\
           -1.25 \\ 
            0.25 
          \end{array} \right), 
        & 
        \bvec{x} = \left(
          \begin{array}{c}
            0.0 \\ 
            0.25  \\
           -0.75
          \end{array} \right) 
        \notag
      \end{eqnarray}
    \item[] Test $2(g)$: Pre-conditioned version $2(g)$ with $\kappa \approx 1$ 
      \begin{eqnarray}
        \boldsymbol{M} = \left( 
          \begin{array}{ccc}
            6.1795   &   11.8207  &   2.0583 \\ 
            15.673   &  -7.56717  &  -3.8520 \\ 
           -5.6457   &   7.96872  &  15.9418 
          \end{array} \right),
        & 
        \bvec{Y} = \left(
          \begin{array}{c}
            1.4114  \\
            0.9972  \\
            9.9643 
          \end{array} \right), 
        & 
        \bvec{x} = \left(
          \begin{array}{c}
            0.0   \\ 
            0.25  \\
           -0.75
          \end{array} \right) 
        \notag
      \end{eqnarray}
    \end{enumerate}
\end{enumerate}

\pagebreak
\begin{acknowledgments}
We received funding for this work from the {\em ASC Beyond Moore’s Law Project} 
at LANL.  We would like to thank Andrew Sornborger, Patrick Coles, Rolando Somma,
and Yi\u{g}it Suba\c{s}\i \, for a number of useful conversations.
\end{acknowledgments}


%


\begin{thebibliography}{99}
\bibskip

\bibitem{hhl}
  A. Harrow, A. Hassidim, and S. Lloyd, 
  Phys. Rev. Lett. {\bf 103}, 150502 (2009). 

\bibitem{ref2013a}
  Stefanie Barz, Ivan Kassal, Martin Ringbauer, Yannick Ole Lipp, Borivoje Dakic ́, 
  \hbox{Ala ́n} Aspuru-Guzik, Philip Walther, 
  {\em  Solving systems of linear equations on a quantum computer},
  arXiv:1302.1210v1 (2013).  

  \bibitem{ref2013b}
  Stefanie Barz, Ivan Kassal, Martin Ringbauer, Yannick Ole Lipp, Borivoje Dakic ́,
   Ala ́n Aspuru-Guzik, Philip Walther, 
  {\em  Solving systems of linear equations on a quantum computer},
  arXiv:1302.1946v1 (2013).

\bibitem{ref2013c}
  X.-D. Cai, C. Weedbrook, Z.-E. Su, M.-C. Chen, Mile Gu, M.-J. Zhu, Li Li, 
  Nai-Le Liu, Chao-Yang Lu, Jian-Wei Pan, 
  {\em  Experimental quantum computing to solve systems of linear equations},
  arXiv:1302.4310v2 (2013).  

\bibitem{fggs}
  Edward Farhi, Jeffrey Goldstone, Sam Gutmann and Michael Sipser,
  {\em Quantum Computation by Adiabatic Evolution},
  arXive:000110 (2000). 

\bibitem{Ising1925}
  E. Ising,  {\em Z. Phys.} {\bf 31} (1925) 253.
 

\bibitem{gray_code}
  Press, William H.; Teukolsky, Saul A.; Vetterling, William T.; Flannery, Brian P. 
  {\em "Section 22.3. Gray Codes". Numerical Recipes: The Art of Scientific Computing (3rd ed.)},
  New York, USA: Cambridge University Press. ISBN 978-0-521-88068-8 (2007).

\bibitem{messiah}
  Messiah, Albert
  {\em ``Chapter XVII.'' Quantum Mechanics.}
  Dover Publications. ISBN 0-486-40924-4. (1999)

\end{thebibliography}
\end{document}